\documentclass[12pt,letterpaper,draft]{article}
\usepackage{amsmath}
\usepackage{amsthm}
\usepackage{amscd}
\usepackage{amsfonts}
\usepackage[psamsfonts]{amssymb}
\usepackage{color}

\newcommand{\be}{\begin{equation}} \newcommand{\ee}{\end{equation}}
\newcommand{\bea}{\begin{eqnarray}} \newcommand{\eea}{\end{eqnarray}}
\newcommand{\nn}{\nonumber} \def\ba{\begin{array}} \def\ea{\end{array}}

\newcommand\belabel[1]{\begin{equation}\label{#1}}
\newcommand\bearlabel[1]{\begin{eqnarray}\label{#1}}
\newcommand\bra[1]{{\langle {#1}|}}  
\newcommand\ket[1]{{|{#1}\rangle}}  

 \def\a{\alpha} \def\b{\beta} \def\g{\gamma}
\def\r{\rho}

\def\IC{\mathbb{C}} \def\IZ{\mathbb{Z}} \def\IR{\mathbb{R}}

\def\NN{\text{NS-NS}}
\def\RR{\text{R-R}}

 \def\half{\frac{1}{2}} \def\wt{\widetilde}
\def\Tw{\text{T}}         
\def\Tr{\mbox{Tr}}

\usepackage{amssymb} \usepackage{latexsym}
\input xypic 

   \def\Z{\mathbb{Z}}
  
  \def\C{\mathbb{C}}
  
  \def\R{\mathbb{R}}
\def\C{\mathbb{C}} \def\c{\mathcal{C}} 
  
  \def\Pi{\mathbb{P}^{\infty}}

\def\Zpk{\mathbb{Z}/p^{k}} \def\Zpk1{\mathbb{Z}/p^{k-1}}

\newcommand{\rref}[1]{(\ref{#1})}

\newcommand{\beg}[2]{\begin{equation}\label{#1}#2\end{equation}}
\def\r{\rightarrow}

\def\sl2{\widetilde{SL_{2}(\Z)}}

\makeatletter \@addtoreset{equation}{section} \makeatother

\begin{document}
\begin{titlepage}


\leftline{\tt hep-th/0703122}

\vskip -.8cm

\rightline{\small{\tt MCTP-07-05}}

\vskip 1.7 cm

\centerline{\bf \Large  Comments on $D$-branes on Orbifolds and K-theory}

\vskip 1.7 cm

\centerline{\large Igor Kriz$^{1}$, Leopoldo A. Pando Zayas$^{2}$ and Norma 
Quiroz$^{3,4}$ }

\vskip 1cm

\centerline{ \it ${}^{1}$Department of Mathematics,  The University of Michigan}
\centerline{\it Ann Arbor, MI 48109-1043, USA}
\vskip .5cm 
\centerline{ \it ${}^{2}$Randall Laboratory of Physics,  The University of Michigan}
\centerline{\it Ann Arbor, MI 48109-1040}
\vskip .5cm
\centerline{\it ${}^{1,2}$ Michigan Center for Theoretical
Physics,  The University of Michigan}
\centerline{\it Ann Arbor, MI 48109-1040, USA}

\vskip .5cm
\centerline{\it ${}^3$ CEFIMAS, Av. Santa Fe 1145}
\centerline{\it 
C1059ABF - Buenos Aires, Argentina}

\vskip .5cm
\centerline{\it ${}^4$ Universidad Nacional de La Plata,}
\centerline{\it C.C.67, (1900) La Plata, Argentina.}

\vspace{1cm}

\begin{abstract}
We systematically revisit the description of $D$-branes on orbifolds and the 
classification of their
charges via K-theory. We include enough details to make the results 
accessible to both
physicists and mathematicians interested in these topics.  
The minimally charged branes predicted by K-theory in $\mathbb{Z}_N$ orbifolds 
with $N$ odd are only BPS. We  confirm this result using the boundary 
state formalism for $\mathbb{Z}_3$. For $\mathbb{Z}_N \times \mathbb{Z}_N$   
orbifolds with and without discrete torsion, we show that the K-theory  classification of charges agrees 
with the boundary state approach, largely developed by Gaberdiel and collaborators, 
including the types of representation on the 
Chan-Paton factors. 
\end{abstract}

\end{titlepage}

\section{Introduction}

Much of the recent progress in string theory revolves around the
concept of Dirichlet-branes. These are objects that source  Ramond-Ramond bosonic massless fields of  
type II and type I string theories carrying one unit of charge \cite{Dai:1989ua,Polchinski:1995mt}.  
More precisely, D-branes are non-perturbative states that enter in the theory as boundaries for the closed string world-sheet, 
introducing  open strings in a consistent way. This definition leads to a  geometrical interpretation of D-branes 
as a surface 
where the ends of open strings are free to move; the ends of the 
open string satisfy Dirichlet boundary conditions along the normal 
directions to the surface, hence their name.  If the surface has $p$ 
dimensions, the brane is denoted shortly as $Dp$-brane. The excitations of the $Dp$-branes 
are described by the open strings which end on it. Branes with the opposite 
charge are called anti-$Dp$-branes (or  $D\bar p$-branes for short).

For some branes supersymmetry plays a crucial role. In particular, some branes are 
states that preserve half of the spacetime supersymmetries, they are known as 
Bogomolnyi-Prasad-Sommerfield (BPS) branes. These $D$-branes are stable objects since
they form short representations of the supersymmetry algebra. Subsequently, their mass  is completely determined by their
charge ($m=|Q|$ in the appropriate units).

These BPS branes constitute key evidence in the formulation of various 
strong/weak  duality conjectures among various string theories.  An analysis of some duality conjectures 
beyond supersymmetry brought about the discovery of new solitonic states in 
string theory.  These states are non-BPS branes as they do not preserve any spacetime supersymmetry and consequently 
the mass and charge relationship is generic ( $m>|Q|$). Interestingly, these states  can be stable if they 
are the lightest states of the theory 
\cite{Sen:1998ii,Sen:1998rg,Sen:1999mg}. 

D-branes figure prominently is string phenomenology. The goal of the string phenomenology program 
is to find $D$-brane 
configurations with world-volume field theories resembling the Standard 
Model of Particles. Some concrete models have been 
proposed based on: Calabi-Yau compactifications, orbifold and orientifold compactifications, 
orbifold singularites like $\IC ^3/\IZ_N$ and $\IC ^3/\IZ_N\times \IZ_N$, 
among others.  At the moment there is not a single model that 
could be considered completely realistic. The search for realistic models requires 
knowledge of the {\it full spectrum of $D$-branes and their conserved charges} in a wide range of  
string theories.

There are two main approaches towards the construction of the {\it  spectrum of D-brane charges}. 
The first one is given by the
boundary state formalism wherein a $D$-brane  is a coherent state in the closed string theory 
that emits or 
absorbs closed string states \cite{DiVecchia:1999rh,DiVecchia:1999fx,Gaberdiel:2000jr}. The 
boundary state has to be a physical state of the closed string theory and therefore must satisfy 
several conditions (see section \ref{sec:boundarystates}). These conditions imply that a $D$-brane is  a 
linear combination of boundary states defined for the different sectors of 
the closed string theory. Crucially, this linear combination of boundary states  encodes 
the information of the open string spectrum characterizing the $D$-brane. The consistent set of solutions 
of the physical conditions  is the way the boundary state formalism answers the question of the 
spectrum of $D$-branes. This construction of 
$D$-branes does not rely on space-time supersymmetry and allows us to analyze 
the spectrum of $D$-branes in backgrounds where supersymmetry is broken or 
does not exist at all, as in type 0A and 0B theories.

Let us mention an important property of the boundary state formalism approach to the 
classification of D-brane charges. 
For a given set of  Ramond-Ramond (R-R) charges, a  $D$-brane configuration 
exist with these charges.  The space of all R-R charges is a lattice.  
The basis vectors of this lattice are charges that corresponds to fundamental 
branes. By fundamental we mean branes that can not be written as linear 
combinations of others, hence carrying smaller charges and masses. Therefore, a 
general charge in the lattice can be generated by  an integer linear 
combination of these fundamental branes. 

The other approach for the classification  $D$-branes, based largely on 
Sen's construction \cite{Sen:1999mg}, is  K-theory.
The  spectrum of open strings with both ends on a $D$-brane contains,
among other states, massless vectors. They are $U(1)$ gauge fields
living on the $D$-brane world-volume.
By considering a set of $N$ coincident branes, one
obtains a
non-Abelian group $G$; for Type II string theory, it is $U(N)$. Therefore,  
$N$ $D$-branes are roughly  described by a $G$ gauge
bundle whose base space is the world-volume of the
branes. In this way, a system of $N$ coincident 
$D$-branes and $N$ coincident anti-$D$-branes is characterized  by a pair of vector
bundles (a $G_1$ gauge bundle
for the $D$-branes and a $G_2$ gauge bundle for the anti-$D$-branes).
This system of branes-antibranes is unstable
since the spectrum of the open string stretched between the branes and the anti-branes contains a tachyon.
For the case in which the two gauge bundles are the same, the system carries 
no net charge and the instability  allows  the
possibility  of brane-antibrane annihilation, recovering the closed string 
vacuum state and preserving charge conservation. This leads to introducing 
equivalence classes of pairs of bundles.
Such classes are naturally elements of K-theory. 
The original suggestion of \cite{Minasian:1997mm} stating that a natural 
framework for $D$-brane charges is K-theory was further tested  in 
\cite{Witten:1998cd} for Type IIB string theory and Type I in ten 
dimensions, the IIA case was discussed in \cite{Horava:1998jy}.

The initial statement that K-theory
describes the complete space of conserved charges was generalized to various 
  settings: string theory on orbifolds, 
orientifolds and backgrounds with a B-field. The relevant K-theories were determined to be 
the equivariant, real and twisted K-theory, respectively. 
The general believe is that: for each string model there exists a generalization of K-theory that 
classifies the D-brane charges. The boundary state formalism allows to compute the spectrum of 
D-brane charges. The space of all conserved charges is a sublattice of the 
full lattice of charges. As we have mentioned above, the fundamental branes 
obtained by boundary states are the  generators of this full lattice.  Then, 
it is enough to construct the fundamental branes in the specific string 
models in order to compare the results with K-theory predictions. Some 
tests of this version of the K-theory conjecture have been realized using the boundary 
state formalism in several simple orbifold and orientifold models 
\cite{Stefanski:2000fp,Quiroz:2001xz,Maiden:2006qe}.

By scrutinizing 
more general physical situations it has become apparent that
ordinary K-theory is not enough to characterize  $D$-brane charges. For 
example, considerations of
string theory in backgrounds with nontrivial torsion classes suggests that 
perhaps
$D$-brane charges are calculated by differential K-theory 
\cite{freedmooresegal}. Considerations of
certain anomalies in IIA lead to the conjecture that the relevant object to 
classify
$D$-brane charges is Elliptic Cohomology \cite{kriz}.
K-theory  also has problems in classifying Ramond-Ramond fluxes. In 
such case, the classification is  incompatible with  S-duality in type IIB 
\cite{Diaconescu:2000wy}. More
generally, it is fair to state that  the generalized cohomology theory that 
completely
characterizes $D$-brane charges in the general case is still not completely
understood.

\subsection{Outline}

In this paper we explicitly present the K-theory group for general 
Abelian orbifolds.  We shall consider the simplest non-trivial examples of 
orbifolds like  $\IC^3/\IZ_N$ and $\IC^3/\IZ_N \times \IZ_N$ and consider  
the inclusion of discrete torsion when appropriate.

Interestingly, an analysis of K-theory suggests an asymmetry in 
the K-theory
of $\mathbb{Z}_N$ orbifolds depending on whether $N$ is even or odd. This 
result motivates us to consider boundary states with the intention of 
verifying this prediction.
In particular, K-theory predicts that the $\mathbb{Z}_3$ orbifold should 
have only BPS branes.  We proceed to verify this prediction
using the boundary state formalism.
We compare the K-theory predictions for the $\IC^3/\IZ_N \times 
\IZ_N$ without and with discrete torsion orbifolds   with the 
results obtained by Gaberdiel and Craps \cite{Craps:2001xw} using boundary 
states. We find full agreement, including the type of representations (conventional or 
projective) acting on the Chan-Paton factors.

Some of the results presented here are known in the literature,
however, we feel that a systematic presentation of the mathematical and 
physical sides
is lacking for orbifolds  like $\IZ_N$ and $\IZ_N \times \IZ_N$ without and  
with discrete torsion. There are various reviews that exhaustively present 
some particular aspects of the topics we touch briefly here. We refer the reader 
interested in  details  in  the construction of boundary states to the 
excellent reviews  \cite{DiVecchia:1999rh,DiVecchia:1999fx,Gaberdiel:2000jr,Craps:2000zr}, and  for
K-theory aspects to \cite{szabo,Moore:2003vf}. Our goal in reviewing part of the
material on perturbative construction of $D$-branes is to allow for
an easy introduction to the interested mathematical reader, likewise we 
present many explicit calculations
of K-theory which should be accessible to interested physicists.

The organization of the paper is a follows. We start with a general discussion 
of $D$-branes. Section \ref{sec:boundarystates} contains a detailed 
description of the boundary
states techniques. Section \ref{sec:orbifolds} discusses strings and branes 
on orbifolds including the case of discrete torsion. 
Section \ref{sec:orbifoldsexamples} contains a number of explicit examples 
including $\mathbb{Z}_N$ with even and odd $N$. This section contains an account of the 
relevant K-theory.
Section \ref{sec:discretetorsion} discusses the presence of discrete torsion 
in various types of orbifolds. We present
some conclusions and point out to some interesting open questions in 
\ref{sec:conclusions}. In two appendices we
collect a number of technical results used in the main text. Appendix  \ref{sec:stringappendix}
contains perturbative string theory  results and appendix \ref{sec:mathappendix} contains 
results of from K-theory .

\section{$D$-branes as boundary states}
\label{sec:boundarystates}
In this section we briefly present the construction of $D$-branes as 
boundary states. We refer the 
reader interested in
the details to the excellent reviews   \cite{DiVecchia:1999rh,DiVecchia:1999fx,Gaberdiel:2000jr,Craps:2000zr}.
We begin with the open  and closed string description of $D$-branes and 
follow with a review of the boundary
state formalism. We will see that to describe $D$-branes, the 
boundary state has to satisfy
various conditions such as invariance under all symmetries of the string
theory  and a  consistent open/closed string interaction. Finally, we review 
the classification
of $D$-branes in Type IIA/IIB in ten dimensions, using the boundary state 
and K-theory formalism.

\subsection{Open  and Closed string description of $D$-branes}
\label{subsec:modular}

We work in the  R-NS formalism and the light-cone  gauge 
quantization. We follow 
the notation  of \cite{Sen:1998ii,Gaberdiel:2000jr}. 
A $Dp$-brane is a hyper plane, in ten dimensional space-time, that 
extend along $p$ spatial directions
and where the endpoints of open strings can end.  In other words,  a 
$Dp$-brane  is defined by the boundary
conditions the endpoints of open strings satisfy. There are Neumann boundary 
conditions along the
$p+1$-directions (including time):
\be
\label{eq:openbcN}      
\partial_{\sigma}
X^\mu(\sigma,\tau)|_{\sigma=0,\pi}  =  0 \;,\qquad  \hbox{$\mu=0,\ldots, p$} 
\\
\ee
and Dirichlet boundary conditions  along $9-p$ directions (the transversal 
directions to the $Dp$-brane):
\be
\label{eq:openbcD}    
X^\nu(\sigma,\tau)|_{\sigma=0,\pi}  =   a^\nu \;,\quad 
\hbox{$\nu=p+1,\ldots,9$}\,,
\ee
where  $X^{\mu}$ are the bosonic world-sheet fields which are 
maps from the world-sheet into spacetime.  The constants $a^{\nu}$ denote the
position of the $Dp$-brane in space-time and the parameters $\sigma$ and $\tau$ are  the spatial and 
temporal coordinates on
the world-sheet respectively. The end points of the strings are at $\sigma = 
0 ,\pi$.  The interaction
between two $D$-branes is
given by vacuum fluctuations of an  open string  beginning on one $D$-brane 
and ending on the other
$D$-brane propagating in a loop with periodic time $\tau \in [0,2\pi t]$. 
Graphically,  the topology
of the open string world-sheet is a cylinder ending on the two branes. The 
interaction
process is described by the one-loop amplitude
\be
\label{eq:oneloop}
\int_0^{\infty} \frac{dt}{2t}\Tr(\hat P e^{-2tH_o})\,,
\ee
where  $t$ is the modulus of the cylinder and runs
over the range $0<t < \infty$. The trace is taken over the open
string spectrum and weighted by the exponential of the Hamiltonian denotes a 
partition
function with  a projector operator  $\hat P$ inserted. We will return to 
the role of 
this operator in some concrete examples. The Hamiltonian of the open string 
$H_0$ is explicitly defined
in appendix \ref{sec:stringappendix}.

Since the theory is conformal invariant, one can always find a conformal 
transformation
such that the world-sheet coordinates are exchanged: $\sigma \leftrightarrow 
\tau$. After
a conformal rescaling of the world-sheet coordinates  the interaction 
between $D$-branes
has the topology of a cylinder  with length parameterized
by $l=1/2t$ \cite{DiVecchia:1999rh,Gaberdiel:2000jr,Polchinski:1998rq}. It 
is drawn by a
closed string state of length $2\pi$ propagating between the branes  in a 
euclidean time $2\pi l$. The ends
of the
cylinder lie on the $D$-branes and represent boundary closed string states 
that are created or annihilated
by the branes.  In this way a $D$-brane is a boundary or a source of closed 
strings. The interaction process
is described by the tree-level closed string amplitude
\be
\label{eq:tree}                    
\int_0^{\infty} dl\, \bra{Dp'}\;e^{-lH_c}\ket{Dp}\;,
\ee
where $H_c$ is the closed string Hamiltonian described in appendix 
\ref{sec:stringappendix}.

These two descriptions of the interaction of $D$-branes are  physically 
different
but they are equivalent in the sense that the interaction amplitudes   are  
related to each other by a conformal transformation. This equivalence is referred to
as open/closed string duality or world-sheet duality.

\subsection{Boundary States}

Under the modular transformation, the open boundary conditions 
(\ref{eq:openbcN}) and (\ref{eq:openbcD}) become
boundary conditions for closed strings states \cite{DiVecchia:1999rh}:
\be\label{eq:closedbc} 
\begin{array}{rcll}
\partial_\tau X^\mu(\sigma,0) |Bp\rangle& = & 0 \qquad &
\hbox{$\mu=0,\ldots, p$} \\
X^\nu(\sigma,0) |Bp\rangle & = &  a^\nu |Bp\rangle \quad &
\hbox{$\nu=p+1,\ldots,9$}\,.
\end{array}
\ee
These boundary conditions are  defined at $\tau=0$  but similar conditions 
are imposed at $\tau=2\pi l$.

In order to solve the equations \eqref{eq:closedbc}, we expand the closed string 
coordinate
operators  $X^{\mu}$ in terms of the oscillator  modes. The light-cone
coordinates  $X^0,X^9$ satisfy Dirichlet boundary conditions 
\cite{Green:1996um}.
In this context we will be describing D-instantons, but performing an 
appropriate Wick rotation one can transform these states back to ordinary 
$D$-branes.
The boundary conditions \eqref{eq:closedbc} become
\be
\label{eq:bosonbc} 
\ba{rcll}
{\hat p}^\mu |Bp\rangle &
= & 0  \qquad & \hbox{$\mu=1,\ldots, p+1$}, \\
\left(\alpha^\mu_n + \widetilde\alpha^\mu_{-n} \right)
|Bp\rangle &
= & 0  \qquad & \hbox{$\mu=1,\ldots, p+1$}, \\
\left(\alpha^\nu_n - \widetilde\alpha^\nu_{-n} \right)
|Bp\rangle &
= & 0 \qquad & \hbox{$\nu=p+2,\ldots,8$}, \\
{\hat x}^\nu |Bp\rangle & = & b^\nu |Bp\rangle \quad &
\hbox{$\nu=0,9,p+2,\ldots,8$}\;, \\
\ea
\ee
where ${\hat p}^{\mu}$ is the center of mass momentum operator,
$\alpha_n$ and $\tilde \alpha_n$ are the left- and right-moving modes
of the bosonic operator $X$ with $n\in\IZ$ and $\hat x^{\nu}$ is the
center of mass position operator. In supersymmetric string theories one has 
to include
analogous boundary conditions for the fermions

\be\label{eq:fermionbc}         
\ba{rcll}
\left(\psi^\mu_r + i \eta \widetilde\psi^\mu_{-r} \right)
|Bp,\eta\rangle &
= & 0  \qquad & \hbox{$\mu=1,\ldots, p+1$}, \\
\left(\psi^\nu_r - i \eta \widetilde\psi^\nu_{-r} \right)
|Bp,\eta\rangle &
= & 0 \qquad & \hbox{$\nu=p+2,\ldots,8$} \,,
\ea
\ee
which define the fermionic part of the boundary state. Here, $\psi_r$ and 
$\tilde \psi_r$ are
the left- and right-moving modes of the fermion operators; $r \in \IZ$ for 
the R sector and $r \in \IZ + \half$ for the NS sector; $\eta=\pm 1$
denotes the spin structure. These boundary conditions can be solved 
separately for the different closed
string sectors of the theory.  Using the techniques of
\cite{Callan:1987px} one can easily find the solution to equations 
(\ref{eq:bosonbc}) and
(\ref{eq:fermionbc}). Since  left and right movers are related at the 
boundaries, only boundary states in the NS-NS and R-R sector are allowed 
\cite{ DiVecchia:1999rh}. The direct product of the solutions of these 
equations determines  a boundary state in the Fock space of the closed 
superstring in the NS-NS and R-R sector

\be
\label{eq:coherents}
\widehat{|Bp, k,\eta\rangle}_{\substack{
\text{NS-NS}\\
\text{R-R}}}
=  \exp\left\{ \sum_{n>0}
\frac{1}{n}
        \alpha^\mu_{-n}S_{\mu \nu} \widetilde\alpha^\nu_{-n}
+ i \eta \sum_{r>0}
\psi^\mu_{-r}S_{\mu \nu} \widetilde\psi^\nu_{-r}
\right\} |Bp,k,\eta\rangle^{(0)}_{\substack{
\text{NS-NS}\\
\text{R-R}
}},
\ee
where $S_{\mu \nu}$ is a diagonal matrix encoding the boundary conditions of 
the D$p$-brane. It has entries equal
to $-1$ for the $p+1$  Neumann  directions and  +1 for the  $7-p$ Dirichlet
boundary conditions. From the boundary conditions  \eqref{eq:bosonbc}, the 
momentum on the Neumann
directions is zero. So, the boundary state carries momentum $k$ only along 
the Dirichlet directions.
The state $|Bp, k,\eta\rangle^{(0)}$
denotes the Fock vacuum. In the NS-NS sector it is the same as the  ground 
state of the
closed string. It is unique and independent of the spin structure. In the 
R-R sector
the Fock vacuum needs special attention since it is defined by the
zero modes and is degenerate. The precise definition of the R-R ground state
is given in appendix \ref{sec:stringappendix}.

It is convenient to work with a localized
boundary state for which one has to take the Fourier transform
of equation  (\ref{eq:coherents})
\be
\label{eq:fourier}
\ket{Bp, a,\eta}_{\substack{
\text{NS-NS}\\
\text{R-R}
}}
= \int
\prod_{\mu=0,9,p+2,\ldots, 8} dk^\mu e^{i k\cdot a}
\widehat{\ket{Bp,k,\eta}}_{\substack{
\text{NS-NS}\\
\text{R-R}
}}\,,
\ee
where $a$ denotes the position vector of the brane in the
Dirichlet directions. This  form of the boundary state is suggested by the 
consistency conditions  described below.

\subsection{Consistency conditions on the Boundary States}
So far we have constructed boundary states as solutions to the closed 
boundary conditions \eqref{eq:closedbc}.  A
$D$-brane will be described by a linear combination of boundary states 
\eqref{eq:fourier} defined in the different
sectors of the closed string  and carrying different spin structures. The 
linear combinations are determined
essentially  by three requirements.   First, a boundary state describing a 
$Dp$-brane has to be a physical state
of the Hilbert space ${\cal H}$ in the  closed string which means that it has to 
be invariant under all symmetries
and projection operators of the theory. In particular, the boundary state has 
to be  GSO-invariant.  If orbifold or
orientifold symmetries are considered, the boundary state should be 
invariant under these symmetries.
Secondly, a $D$-brane constructed by boundary states should contain all the 
information about the open string
defining the $Dp$-brane. The  open string spectrum can be determined by 
computing the interaction amplitude
between two $Dp$-branes and  after a conformal transformation it should be 
expressed into a one-loop open string amplitude
\be
\label{eq:treeopen}
\int_0^{\infty} dl \bra{Dp}\;e^{-lH_c}\ket{Dp}=\int_0^{\infty} 
\frac{dt}{2t}\Tr(\hat P e^{-2tH_o})\,,
\ee
that is, the $Dp$-brane defined by boundary states should satisfy the 
open/closed string duality.
Third, the open string introduced in this way  must have consistent 
interactions  with the
closed string sector of the theory. It means that the end points of the open 
string lying on
the $D$-brane should be able to join to form a physical closed string state 
of the theory.

These conditions are intrinsic of an interactive string theory  and they do 
not rely in space-time supersymmetry. It allows one to construct branes in 
supersymmetric  and non-supersymmetric
string theories as well as supersymetric (BPS) and non-supersymmetric 
(non-BPS) branes \cite{Bergman:1999km,Bergman:1999kz}.

We have mentioned in Sec.\ref{subsec:modular} that a $D$-brane has a 
geometrical interpretation
as an hyperplane where open strings can end and we have seen that 
consistency conditions  force
the boundary states to encode all the information about these open strings.  
In this sense, the
interpretation of $D$-branes as boundary states described above relies in 
the space-time geometry. In a
general context, one could be interested in analyzing $D$-branes in string 
theories without referring
to the geometry of the space-time. There are several examples  of string 
theories, for instance,
Gepner models, WZW models, two-dimensional string theories, where space-time 
is partially replaced
by a conformal field theory. In  these cases the formalism of boundary 
states is more powerful.

Generically, a boundary state in a rational conformal field theory is a 
boundary that satisfy
the gluing conditions $(W_n -\wt W_{-n})\ket{i}=0$ where $W$ and $\wt W$ are 
the
generator of the symmetry algebra of the theory.  It means that the boundary 
state preserves a
diagonal symmetry algebra. Solutions to these equations are called  
Ishibashi
states \cite{Ishibashi:1988kg,Onogi:1988qk}.
The boundary state is a linear combination of Ishibashi states:
$\ket{\alpha}= \sum_j B^j_{\alpha}\ket{j}$ where the coefficients $B^j$  are 
restricted
by the Cardy condition, i.e.; by the modular
transformation of the partition functions. This condition is
equivalent to the open-closed string duality.
If there are more symmetry algebras in the system the boundary also has to  be 
a solution
of the gluing conditions defined by the generators of such
algebras \cite{Recknagel:1997sb,Gaberdiel:2002my}. The Cardy condition 
produces  $D$-branes as
boundary states that are acceptable from the perspective of CFT. But when 
applying it to
string theories, the Cardy condition is not enough to produce $D$-branes, 
since
string theories include other physical considerations.

\subsection{Review of $D$-branes in ten dimensions}
\label{subsec:10d}

{\it BPS branes}\\
As an example of $Dp$-branes  described by boundary states we review the 
simplest
case of $D$-branes in IIA or IIB in ten dimensions. In these theories there 
is only a projector operator, the GSO. So, the first task is to see if the
the boundary state defined by equations \eqref{eq:coherents} and 
\eqref{eq:fourier} is GSO invariant. In the  NS-NS, the ground state is 
taken to have negative eigenvalue under the action of the GSO operators 
$(-1)^F$ and $(-1)^{\wt F}$. The world-sheet fermions $\psi^{\mu}$ and $\wt 
\psi^{\mu}$
anti-commute with these operators. From these facts, it is easy to prove 
that

\begin{align}
(-1)^F\ket{Bp,a,\eta}_{\text{NS-NS}}&= -\ket{Bp,a,-\eta}_{\text{NS-NS}}\;,
&(-1)^{\wt F}\ket{Bp,a,\eta}_{\text{NS-NS}}&= 
-\ket{Bp,a,-\eta}_{\substack{\text{NS-NS}}}\,.
\end{align}
The GSO operator change the spin structure of the boundary states. Therefore, 
the linear combination  of boundary states in the NS-NS sector with opposite 
spin structure will be GSO invariant

\be
\label{eq:nsnsbs}
\ket{Bp, a}_{\text{NS-NS}}=\frac{1}{2}(\ket{Bp, 
a,\eta}_{\text{NS-NS}}-\ket{Bp, a,-\eta}_{\text{NS-NS}})\;.
\ee

In  the R-R sector the GSO  operator has the form
$(1+(-1)^F)(1\pm (-1)^{\bar F})$, with the positive sign corresponding to 
Type IIB and the negative  to Type IIA. From equation \eqref{eq:gsogs} and 
\eqref{eq:coherents} the action of $(-1)^F$ and $(-1)^{\wt F}$  on the R-R 
boundary state is
\bea
\label{eq:gso}
(-1)^F |Bp,{\bf a},\eta\rangle_{\RR}
& = & |Bp,{\bf a},-\eta\rangle_{\RR} \,, \nonumber\\
(-1)^{\widetilde{F}} |Bp,{\bf a},\eta\rangle_{\RR}
& = & (-1)^{p+1} |Bp,{\bf a},-\eta\rangle_{\RR} \,,
\eea
and the  GSO-invariant state in the R-R sector is
\be
\label{eq:rrbs}
\ket{Bp,a}_{\text{RR}}= 
\frac{4i}{2}(\ket{Bp,a,\eta}_{\RR}+\ket{Bp,a,-\eta}_{\RR})\;.
\ee
with $p$ even for IIA and $p$ odd for IIB.

Although the boundary states \eqref{eq:nsnsbs} and \eqref{eq:rrbs} are GSO 
invariant, one has to take a linear combination of these states in order to 
define a $D$-brane
\be
\label{eq:bsbps}
\ket{Dp, a}={\cal N} (\ket{Bp,a}_{\NN}+ \epsilon \ket{Bp,a}_{\RR})\;.
\ee
The parameter $\epsilon=\pm 1$ describes the R-R charge of the brane. It can 
be determined by saturating
the boundary state with the corresponding R-R vertex operator 
\cite{DiVecchia:1997pr}. By convention, the positive sign corresponds
to a $Dp$-brane and the negative sign to an anti-$Dp$-brane.
The normalization constant ${\cal N}$ is to be determined.

This linear combination, as well as the coefficients defining the NS-NS and 
R-R boundary states, is  suggested by the conditions that the tree-level 
amplitude of the $D$-branes has to be equivalent to a one-loop open string 
amplitude.  To see that, one replaces  \eqref{eq:nsnsbs}, \eqref{eq:rrbs}  
and \eqref{eq:bsbps}
into \eqref{eq:tree}  and after the transformation $l=1/2t$ one obtains  
that each component of the $D$-brane produces an open string partition 
function

\begin{align}
\label{eq:amplitudes}
\int_{0}^{\infty} dl \;_{\NN}\bra{Bp, a,\eta}e^{-lH_c}\ket{Bp, a,\eta}_{\NN} 
&=
\int_{0}^{\infty} \;\frac{dt}{2t}\Tr_{\text{NS}} e^{-tH_0}\;,\nn\\
\int_{0}^{\infty} dl\; 
_{\NN}\bra{Bp,a,\eta}e^{-lH_c}\ket{Bp,a,-\eta}_{\NN}&=
\int_{0}^{\infty}\; \frac{dt}{2t}\Tr_{\text{R}}e^{-tH_0}\;.\nn\\
\int_{0}^{\infty} dl 
\;_{\text{RR}}\bra{Bp,a,\eta}e^{-lH_c}\ket{Bp,a,\eta}_{\text{RR}} &=
\int_{0}^{\infty}\; \frac{dt}{2t}\Tr_{\text{NS}}(-1)^F e^{-tH_0}\;,\\
\int_{0}^{\infty} dl\; 
_{\text{RR}}\bra{Bp,a,\eta}e^{-lH_c}\ket{Bp,a,-\eta}_{\text{RR}}&=
\int_{0}^{\infty}\; \frac{dt}{2t}\Tr_{\text{R}}(-1)^Fe^{-tH_0}=0\nn\;.
\end{align}
The right-hand side of  these equations are open string amplitudes and they 
can be expressed in terms of the Jacobi theta functions 
\cite{Polchinski:1998rq}. Expanding the functions around $t \rightarrow \infty$, one can see 
that  the first and third  open string amplitudes given by the right-hand 
side of equation \eqref{eq:amplitudes} contain a tachyon in the spectrum. 
This result will be important when analyzing the stability of the 
$D$-branes.
The equalities in \eqref{eq:amplitudes} are satisfied if the normalization 
constant is
\be
{\cal N}^2 = \frac{V_{p+1}}{(2\pi)^{p+1}}\frac{1}{32}
\ee
where $V_{p+1}$ is the world-volume of the brane which together with
the factor $(2 \pi)^{-(p+1)}$ come from the momentum integration of the open 
string partition function.  We recall that the boundary state 
\eqref{eq:fourier} is the correct one that allows to set the equalities of 
these amplitudes.

One can read from these relations that the tree-level amplitude satisfy the 
open/closed string equivalence
\be
\label{eq:closeopen}
\int_{0}^{\infty} dl \bra{Dp,a}e^{-lH_c}\ket{Dp,a}= 
\int_{0}^{\infty}\frac{dt}{2t} \Tr_{\text{NS-R}} \frac{1+(-1)^F}{2}e^{-tH_0}
\ee
where $\Tr_{\text{NS-R}}$ denotes the difference between the traces in the 
NS and R sector of the open string.  The projector operator
${\hat P}= \frac{1+(-1)^F}{2}$ inserted in the trace is the GSO operator.  
The open string  tachyon is projected out giving rise to a stable brane. 
Since the open string spectrum is supersymmetric, this brane is BPS.

{\it Non-BPS branes}\\
Let us now  consider the   $Dp-D{\bar p}$ system that, in terms of 
boundary states,  is represented by
\be
\label{eq:bsbab}
\ket{Dp,a}+\ket{D\bar p,a}=2\ket{Bp,a}_{\NN},
\ee
where we have used \eqref{eq:bsbps} to obtain the right-hand side up to some 
normalization constant. The combination
of brane-anti-brane breaks  supersymmetry, therefore it is a non-BPS brane. From 
the first and second equations
in \eqref{eq:amplitudes} it follows that the respective open string amplitude has a $\Tr_{\text{NS-R}}$ with 
projector operator $\hat P=1$. Therefore
the tachyon  in the open string spectrum, producing the instability
of the $D$-brane, is preserved. To understand this fact let us write the 
tree-level amplitude of this brane

\begin{align}
\bra{Dp+D\bar p}e^{-lH_c}\ket{Dp+D\bar p}&= 
\bra{Dp,a}e^{-lH_c}\ket{Dp,a}+\bra{Dp,a}e^{-lH_c}\ket{D \bar p,a}\nn\\
&+ \bra{D\bar p,a}e^{-lH_c}\ket{Dp,a}+ \bra{D\bar p,a}e^{-lH_c}\ket{D\bar 
p,a}
\end{align}
the first and last terms are of the form \eqref{eq:closeopen} and represent  
an open string with both ends
lying on a $Dp$-brane  and an open string with both ends lying on an 
anti-$Dp$-brane, respectively. The second
and third term give rise to a $\Tr_{\text{NS-R}}$ with the projector 
operator $\hat P=\frac{1-(-1)^F}{2}$.  In
this case the GSO projection  is opposite to that of equation 
\eqref{eq:closeopen}. The open string amplitude
corresponds to an open string beginning on a $Dp$-brane and ending on and 
anti-$D$-brane and vice versa. The projector
preserves the tachyon present in the NS sector  and it makes the system  
unstable.  The brane \eqref{eq:bsbab} is
not independent as it relies directly on the existence of the BPS brane 
\eqref{eq:bsbps}.

However, inspired
by this construction one can propose and independent brane by
\be
\label{eq:bsnbps}
{\ket{\widehat{Dp'},a}} ={\cal \widehat N}^2 \ket{Bp',a}_{\NN}.
\ee
This brane  produces an open string partition 
function with a NS and R sector unprojected. Therefore
it has a tachyon in the open string spectrum and is unstable. From the same 
reason, the open string spectrum is
non-supersymmetric. The brane is non-BPS. It is an independent brane if $p'$ 
is odd in IIA or  even in Type IIB.
The normalization constant is
\be
{\cal \widehat N}^2 =\frac{V_{p'+1}}{(2\pi)^{p'+1}}\frac{1}{16}.
\ee

The boundary states tell us how the branes couple to the R-R fields of the 
theory, then
we have a way of classifying branes according to the R-R charges. In the 
case in question there is only one kind of
brane carrying a  R-R charge.
This result is in agreement with that given  by classifying the R-R charges 
of the
branes using K-theory \cite{Witten:1998cd}. In this case the analysis is 
given by the K-theory of the transversal
space  $S^n$ to the D$p$-brane, with $9=p+n$. We summarize the results in 
the following tables

\begin{table}[here]
\centering
\begin{tabular}{|c|c|c|c|}
\hline
Brane &  Charges& Type IIA & Type IIB \\ \hline
Stable & 1 & $p$ even & $p$ odd\\
Unstable& 0 & $p$ odd & $p$ even\\
\hline
\end{tabular}
\caption{Classification of R-R charges using boundary states}
\label{table:bsbranes}
\end{table}
\begin{table}[here]
\centering
\begin{tabular}{|c|c|c|}
\hline
       K-theory          &  n even  & n odd  \\ \hline
$\tilde{K}(S^n)$ in Type IIB  & $ \IZ$ &   0\\
  $K^{-1}(S^n)$ in Type IIA& 0 & $\IZ $\\
\hline
\end{tabular}
\caption{Classification of R-R charges by K-theory}
\label{tab:kbranes}
\end{table}

Here $\tilde{K}^i(S^n)$ is the reduced $i$-th $K$-theory
group of the $n$-dimensional sphere; reduced means that
we factor out the corresponding $K$-theory group of a point.
The group only depends on whether $i$ is even or odd.
$\tilde{K}^{odd}=K^{odd}$, since $K^{odd}(*)=0$ where
$*$ denotes a single point.

We have presented shortly the methodology of boundary states
to construct branes in string in ten dimensions but the construction
of boundary states on other string vacuums is similar.


\section{Strings and branes on orbifolds}
\label{sec:orbifolds}
In this section we  briefly review  strings and $D$-branes on orbifolds. 
This formalism will be used in the following chapters. We will first give a physical definition of orbifolds and 
then we will see the conditions
$D$-branes  must satisfy to exist in these theories. Much of the 
presentation conforms to the previous
works \cite{Dixon:1985jw,Dixon:1986jc, 
Dixon:1986yc,Douglas:1998xa,Douglas:1999hq}. We also give an
introduction to orbifolds with discrete torsion 
\cite{Vafa:1986wx,Vafa:1994rv,Aspinwall:2000xv,Gomis:2000ej}. Finally,
we introduce the mathematical ingredients needed for the K-theoretic 
computations.

\subsection{ Orbifold theories}

Mathematically, an orbifold is locally a quotient of a manifold by a finite 
group. We will study special kinds
of orbifolds which are quotients of Euclidean spaces  (for example, 
$\mathbb{R}^6$) by finite groups
which act linearly (real vector spaces on which the
group acts by linear isomorphisms will also be called
real representations).
One can also take a torus $\mathbb{R}^6/L$ where $L$ is some
six-dimensional lattice and other quotient by a finite group acting linearly 
on that space,
which is equivalent to a finite group $G$ on $\mathbb{R}^6$ which preserves 
the lattice $L$.
Clearly, this imposes an additional restriction on the group $G$. 
Additionally, in this
paper, we will only discuss the case of $G$ Abelian ($\mathbb{Z}_N$ and 
$\mathbb{Z}_N\times \mathbb{Z}_N$).

An orbifold ${\cal O}$ is defined  as a space
which is locally the  quotient of a manifold $M$
by the  action of a discrete group $\Gamma$ with finite
stabilizers. In the present paper, we shall only
discuss examples which are {\em global} quotients,
i.e. ${\cal
O}=M/\Gamma$. The discrete  group should be symmetry preserving with
respect the metric\footnote{In the example of lattices
defining  toroidal compactifications,
the orbifold group  has to preserve the inner product between the
vectors basis of the lattice. That is, the action of the orbifold on
any vector of the lattice, has to be an element of the lattice.} of $M$.
In the construction of ${\cal O}$  one has to identify the point $X \in
M$ with all points $hX$, with  $h \in \Gamma$.
The  orbifold may fail to be a manifold at points with non-trivial
stabilizer subgroups,
but it is possible to repair these singularities by removing these  fixed
points and replacing them with a smooth non-compact manifold with
appropriate asymptotic  behavior. This is called  {\it blowing up} or
{\it resolving the singularities}.

However, preserving the singularities is not a problem, since one can
  still have consistent strings propagating on orbifolds
  \cite{Dixon:1985jw,Dixon:1986jc,Dixon:1986yc}. Because points on
  $M$ are identified under elements of the discrete group $\Gamma$, a
  string closes only  up to  an element of $\Gamma$. It means that the
  string fields should satisfy  the boundary condition \be
\label{eq:twisted}
X(\sigma + 2\pi,\tau)= hX(\sigma,\tau)\;.  \ee
This requirement will
factorize the Hilbert space into subspaces\footnote{The decomposition of the 
Hilbert space is into
subspaces  characterized by conjugacy classes; for Abelian groups they
coincide.}  ${\cal H}_h$ for  each $h \in
\Gamma$.   Each sector ${\cal H}_h$ is the Hilbert space for strings
twisted by $h$.  For any non-trivial element $h$, such sectors are
referred to as ${\it twisted}$   ${\it sectors}$ and these states are
only closed on the orbifold space ${\cal O}$. The center of mass of
twisted sector string states is located at the fixed points while  the
oscillators obtain fractional quantum numbers according to the order
of $\Gamma$.

For the identity element of $\Gamma$ the boundary condition for the string 
field  is
$$X(\sigma + 2\pi,\tau)= X(\sigma,\tau).$$
The Hilbert space in this case is referred to as the {\it untwisted
  sector} and the string closes on the original space $M$.

In each subspace ${\cal H}_h$ one has to project onto those states which are 
invariant under the action of $g \in \Gamma$.
For an Abelian group $\Gamma$, the partition function of the orbifold theory 
  is given by

\be
\label{eq:abelianpf}
Z=\sum_h \Tr_{{\cal H}_h} \hat P q^{L_0}\bar q^{\bar L_0},
\ee
where the trace is taken on each sector twisted by $h$, $q=e^{2\pi i 
{\mathfrak t}}$ with $\mathfrak t$ a
complex parameter. The projection operator onto the group invariant
states is $\hat P= \frac{1}{|\Gamma|}\sum_{g \in \Gamma} g$.
One can write this partition function as $Z=\frac{1}{|\Gamma|}\sum_{g,h \in 
\Gamma}Z(g,h)$ with
\be
\label{eq:pf}
Z(g,h)=\Tr_{{\cal H}_h} g q^{L_0}\bar q^{\bar L_0}.
\ee
From the point of view of the two-dimensional field theory on the world-sheet, $Z(g,h)$ represents
the  amplitude in the sector ${\cal H}_h$ for a string to propagate in an 
Euclidean time $ 2 \pi
{\mbox Im } \mathfrak t$ and being translated in $\sigma$ by $2 \pi {\mbox 
Re}
\mathfrak t$ \footnote{The complex parameter $\mathfrak t$ is usually 
denoted by $\tau$ in the
literature, we have chosen this notation to avoid confusion
with the temporal parameter of the world-sheet.}.

It is possible to interpret
$Z(g,h)$  as the partition function over a world-sheet  torus
with modular parameter $\mathfrak t$  and boundary conditions on the string 
coordinates
twisted by the group elements $g$ and $h$ in the $\tau$ and $\sigma$
directions respectively: $X(\sigma + 2\pi,\tau)= hX(\sigma,\tau)\;,
X(\sigma,\tau + 2\pi)= g X(\sigma,\tau)$.

Since the partition function (\ref{eq:pf}) is equivalent to a partition
function of a world-sheet torus, it is easy to see that twisted sectors can 
be obtained  from the
untwisted sector by modular transformations. Then the twisted sectors are 
required by modular
invariance of the torus partition function.

\subsection{Branes in orbifolds and Representations}

Given a string theory on a space $M$, one can analyze the behavior of the
string on the orbifold quotient space ${\cal O}$ following the discussion
above. If the original string theory  on $M$ contains $D$-branes, it will be
interesting to study the action of the orbifold group on the system of such
$D$-branes. In particular, we would like to understand the action of the 
orbifold
group $\Gamma$ on the gauge theory living on the world-volume of the system 
of
$D$-branes. There are two choices one must make. To take a representation  
of
$\Gamma$ on the  space-time, $R(g)$ and  the action of $\Gamma$ on the
Chan-Paton factors $\g(g )$ \cite{Douglas:1996sw} (expanded in 
\cite{johnsonmyers}). In this
way the action of $\Gamma$ on the
fields $\phi= (A_{\mu}(x),X^i(x))$ living on the brane is
\be
\gamma(g)^{-1}\phi \gamma(g) = R(g)\phi\;.
\ee
Invariant states under these projections give the  gauge theory of the 
$D$-branes on orbifolds. From the point of view of open strings, there are as many 
branes as  representations of $\Gamma$
on the Chan-Paton indices. The various $D$-branes differ physically in their 
R-R charges. For instance, the
regular representation corresponds to $D$-branes with untwisted R-R charge. 
Such branes are
called  {\it bulk} branes since they can be localized out of the fixed 
points of the orbifold and
can move freely on the orbifold space. Irreducible representations 
correspond to $D$-branes
carrying   untwisted  and twisted R-R charges. These branes are called
{\it fractional} branes \cite{Diaconescu:1997br} since they carry only a 
fraction of
the charge with respect the the untwisted RR field of a bulk brane and
they are stuck at the orbifold fixed points.
In terms of boundary states, the bulk brane contains only boundary states in 
the untwisted R-R
sector of the closed string. Fractional branes contain both untwisted and 
twisted boundary states.
These two kinds of branes are BPS branes. There is also another  kind of 
branes
in orbifold theories which are non-BPS. In \cite{Gaberdiel:1999ch} they were 
termed truncated branes as they could be seen as a cut off of the fractional 
branes.

\subsection{Discrete torsion}
In the search for more general solutions to string theory satisfying modular
invariance, Vafa \cite{Vafa:1986wx}
realized that it is possible to introduce a phase multiplying the different 
terms in the
partition function of string theory on orbifolds. The name
of discrete torsion refers to
the phase $\epsilon(g,h)$ introduced to define a new orbifold
theory. Modular invariance and higher loop factorization naturally restrict 
the form of the phases.
The freedom of introducing this phase can, in certain cases, be directly 
related to the
B-field.
In \cite{Vafa:1994rv} Vafa and Witten discussed some geometrical 
implications of introducing
nontrivial phases to weight differently certain terms in the string 
partition function.
Various aspects of the spacetime implications of discrete torsion have been 
considered in the literature.
In particular, the effects of discrete torsion on the world volume theory of 
$D$-branes have been discussed
by Douglas and others \cite{Douglas:1998xa,Douglas:1999hq,Aspinwall:2000xv}.

The modular invariant partition function on the torus, mentioned above 
\cite{Vafa:1986wx}, can be written as

\be
Z'(g,h)=\sum_{g,h\in \Gamma} \epsilon(g,h)Z(g,h)\;,
\ee
where $\epsilon(g,h)$ are phases called discrete torsion.
The possible inequivalent phases are determined from
modular invariance  and factorization at higher loops. They should satisfy
\begin{align}
\label{eq:cdt}
\epsilon(gh,k) &= \epsilon(g,k)\epsilon(h,k),\nn\\
\epsilon(g,h) &=\epsilon(h,g)^{-1},\\
\epsilon(g,g)&=1 \nn,
\end{align}
for $g,h,k \in \Gamma$.

These results can be interpreted in terms of group cohomology.
The inequivalent different torsion theories are classified by the second
cohomology group  of $\Gamma$ with values in $U(1)$, $H^2(\Gamma, U(1))$
\cite{Vafa:1994rv}. This cohomology group consists of the two-cocyles 
$c(g,h)\in U(1)$
satisfying the cocycle condition
\be
c(g_1,g_2g_3)c(g_2,g_3)=c(g_1g_2,g_3)c(g_1,g_2)\;.
\ee

Equivalence classes are constructed by the equivalence relation
\be
c'(g,h)= \frac{c_g c_h}{c_{gh}}c(g,h).
\ee
where  $c_g$ and $c_h$ are phases, that is,  $c_g \in U(1)$ for  $g \in  
\Gamma$.  Defining
\be
\label{eq:dt}
\epsilon(g,h)= \frac{c(g,h)}{c(h,g)},
\ee
this discrete torsion phase is the same for  cocycles in the same conjugacy 
class.
The equivalence  classes of cocycles of $\Gamma$ are determined by the second 
cohomology group $H^2(\Gamma, U(1))$.

Discrete torsion can be implemented into the  gauge theory of $D$-branes on 
orbifolds by
requiring  $\gamma(g)$ (the representation of $\Gamma$ on the Chan-Paton 
indices) to
be a projective representation such that

\be
\gamma(g)\gamma(h)=c(g,h)\gamma(gh)\;,
\ee
in this way, any projective representation determines a two-cocycle and in 
this sense we are
incorporating enough information to describe discrete torsion. Recall that the conventional 
representation is defined as $\gamma(g)\gamma(h)= \gamma(gh)$

We will concentrate on orbifolds of the type  $\Gamma=\IZ_n \times \IZ_n$,
the simplest group such that
$H^2(\Gamma, U(1))$ is not trivial. The
generators of this group are $g_1$ and $g_2$. A generic
element of this discrete groups is of the form $g_1^{a}g_2^{b}$ and  will be denoted by $(a,b)$.

The 2-cocyle classes of $H^2(\IZ_n \times \IZ_n,U(1))\cong \IZ_n$.
are represented by
\begin{align}
\label{eq:cocycle}
c^m(g,h): & \;\;\;\;\;\; \IZ_n \times \IZ_n \;\;\; \rightarrow  U(1)\nn\\
           &    ((a,b),(a',b'))                \rightarrow  
\zeta^{m(ab'-a'b)}
\end{align}
where $m=0,1,\ldots, n-1$ denotes the $m$ different elements
of $\IZ_n$ and $\zeta=e^{(2\pi i /n)}$ for $n$ odd, $\zeta=e^{(\pi i /n)}$ 
for $n$ even. The discrete
torsion phase can be obtained by replacing the cocycle (\ref{eq:cocycle}) 
into (\ref{eq:dt}). Then, 
\be
\epsilon(g,h)=\zeta^{2 m (ab'-a'b)}.
\ee
The theory without discrete torsion correspond to $m=0$. Minimal discrete 
torsion corresponds
to the case when $(m,n)=1$, it means, that $\zeta^{2m}$ is a primitive n-th 
root of unity. In
this case, the group $\IZ_n \times \IZ_n$ has a unique projective
representation \cite{Douglas:1999hq,Karpilovsky:1985}.

\section{K-theory for charges of  $D$-branes  on flat $\IZ_N$ orbifolds} 
\label{sec:orbifoldsexamples}

In this section we  begin our  $K$-theoretical description
for groups of charges of $D$-branes in type IIA/IIB string theories
on orbifolds $\mathfrak{O}$. The K-theory groups and boundary states for 
some simple classes of orbifolds in Type II string theories where analyzed 
in \cite{Gaberdiel:1999ch}. We revisit
the boundary states for the non-compactified orbifold $\IZ_2$ presented in 
this work and we express the K-theory results in our notation in order to be 
consistent with the next section.   Later we shall study the K-theory for 
orbifolds of flat spacetime by linear action by the finite abelian groups 
$\IZ_N$, leaving other orbifold models for the next section. In order to 
corroborate our results  for $N$ odd we consider the special case of 
orbifold $\IZ_3$ and construct the boundary states for this model.

\subsection{Generalities of K-theory on orbifolds}
\label{subsec:kth}

We will think of spacetime as
\beg{es1}{V\times \R^{1,1}}
where $\R^{1,1}$ is Minkowski $2$-space (where the light-cone coordinates 
are defined) and $V$ is a {\em complex}
representation of $G$ of complex dimension $4$. The representation
must be complex to preserve supersymmetry, i.e. the reason
for this is physical. $G$ acts trivially on
$\R^{1,1}$.
We have in general the orbifold
$$\mathfrak{O}=(V\times \R^{1,1})/G.$$
The $D$-branes $M$ in \rref{es1} whose images in $\mathfrak{O}$
we will consider will be {\em real} subrepresentations of
$V\times \R^{1,1}$.
This generality is needed to account for all
the branes we consider.
The relevant $K$-groups of interest  are
the equivariant $K$-groups with compact support
\beg{es2}{K^{i,c}_{G}(M^\perp),
}
where $M^\perp$ is the orthogonal complement of $M$ in $V\times \R^{1,1}$,
and $i=1$ or $0$ depending on whether we are in IIA or IIB.

\vspace{3mm}
A basic feature of equivariant $K$-theory is Bott periodicity, which
asserts that for any $G$-space $X$,
\beg{es3a}{K_{G}^{i,c}(X\times W)\cong K_{G}^{i,c}(X)
}
for any {\em complex} $G$-representation $W$.
Now a sum of two copies of any real representation
has complex structure and conversely, an irreducible (hence, since $G$
is abelian, $1$-dimensional) complex representation can be either
irreducible also as an underlying real representation, or can
be a sum of two copies of the same irreducible real representation.
It already follows from this that all that matters for the $K$-group
are the numbers of copies of the individual $1$-dimensional
irreducible real representations on $M$ $\mod 2$, so
we always have
\beg{es3}{K_{G}^{i,c}(M^\perp)\cong K_{G}^{i,c}(M).
}
Additionally, when $M$ itself is a {\em complex} representation of $G$,
then we already know that 
\cite{ascomp}
\beg{es4}{{
\begin{array}{lll}
K_{G}^{i,c}(M)\cong K_{G}^{i,c}(*)& =  R(G)=\Z^{|G|}
&\text{for $i$ even}\\
& =0 &\text{for $i$ odd}
\end{array}
}}
The same analysis holds when the world volume
of the brane $M$ is a sum of an $m$-dimensional
complex $G$-representation and additional $\ell$ real
dimensions on which $G$ acts trivially (assuming
the time-like dimension is included, then $M$
is a $2m+\ell-1$-brane). The only difference between
this case and \rref{es4} is that then the dimensions $i$
have to be replaced by $i+\ell$. In physical language, then,
the group of charges in this case is $\Z^{|G|}$
when $\ell$ is odd in IIA and even in IIB, and $0$ otherwise.

In boundary states, if we denote a $Dp$-brane by
$p=r+s'$, with r the number of direction tangential to the brane
where the orbifold acts trivially and $s'$ the number of Neumann directions
transformed by the group, then
$r\equiv \ell -1$ and $s'\equiv 2m$.

In the case of $|G|$ even, we may encounter real representations.
A general calculation of the groups \rref{es2} was given by
Max Karoubi \cite{karoubi} (see also \cite{hkk}).
All the groups we will need in this paper however can
be calculated from first principles by elementary means.

\subsection{Branes in $\IZ_N$ orbifolds with $N$ even}
\label{subsec:kthNeven}

Let us now see the simple case when $G=\Z/N$ where $N$ is even. 
Then there exists a unique onto homomorphism
$$\phi:\Z/N\r\Z/2,$$
and therefore $\Z/N$ has a unique non-trivial $1$-dimensional
real representation $\alpha$, obtained by composing the sign representation
of $\Z/2$ with the map $\phi$. Suppose the brane $M$ is
a product of an $m$-dimensional complex representation
of $\Z/N$, $s$ copies of the representation $\alpha$,
and $\ell$ other dimensions on which $\Z/N$ acts trivially
(when the time-like dimension is included, it is therefore
a $(2m+s+\ell-1)$-brane). This case reduces to the previous
case when $s$ is even, so let us assume $s$ is odd.

Before proceeding with our K-theory discussion it is necessary to set  the 
notations used in
the mathematical and physical language. A complex coordinate on $\IC^3$ is 
defined by the real
coordinates  $(x^{2i+1},x^{2i+2})$. Then  the 1-dimensional real
representation corresponds to the case when the couple of real coordinates
have mixed boundary conditions, i.e., Neumann-Dirichlet(ND) or DN. The 
complex
representation will corresponds to the case when both coordinates have the 
same boundary conditions, NN or DD. We will be shifting from one notation to 
the other along these notes.

The key observation is that we have a cofibration
sequence
\beg{es4a}{\protect\diagram(\Z/N)/(\Z/(N/2))_+\rto^(.7)f &S^0\rto & 
S^\alpha.
\enddiagram
}

The map $f$ is the only non-trivial (non-constant) map, and
the fact that we have a cofibration sequence \rref{es4a} is
readily verified by definition. Thus, from \rref{esc2}, we obtain
a long exact sequence
\beg{es5}{\diagram \tilde{K}_{\Z/N}^{i-1}(X\wedge (\Z/N)/(\Z/(N/2))_+)
\rto & \tilde{K}_{\Z/N}^{i}(X\wedge S^\alpha)\rto &
\tilde{K}_{\Z/N}^{i}(X)\dto^{f^{*}}\\& & \tilde{K}_{\Z/N}^{i}
(X\wedge (\Z/N)/(\Z/(N/2))_+).
\enddiagram
}


\rref{es5} becomes
\beg{es5a}{\diagram \tilde{K}_{\Z/(N/2)}^{i-1}(X)
\rto & \tilde{K}_{\Z/N}^{i}(X\wedge S^\alpha)\rto &
\tilde{K}_{\Z/N}^{i}(X)\rto^{f^{*}} & \tilde{K}_{\Z/(N/2)}^{i}
(X).
\enddiagram
}
In the case $X=S^0$, $f^*$ in $K^0$ is the reduction map
\beg{es6}{R(\Z/N)\r R(\Z/(N/2))
}
induced by the inclusion $\Z/(N/2)\subset \Z/N$.
(We denote the complex representation ring of $G$ by $R(G)$,
the real representation ring by $RO(G)$).
It then follows that \rref{es6} is an onto map
\beg{es7}{\diagram\Z^{N}\rto^{f^*} & \Z^{(N/2)}.\enddiagram
}
In more detail, the $N$ summands correspond to irreducible
representations of $\Z/N$, and \rref{es7} restricts
the representation to $\Z/(N/2)$. Thus, the kernel is
generated freely by elements of the form
$$\gamma-\gamma\alpha$$
where $\gamma$ is an irreducible (complex) representation of
$\Z/N$. This group is isomorphic to $\Z^{N/2}$. Thus, we have
computed for our brane $M$ in case of $s$ odd
\beg{es8}{{\begin{array}{l}
K^{\ell,c}_{\Z/N}(M)=\Z^{N/2},\\
K^{\ell+1,c}_{\Z/N}(M)=0.
\end{array}
}}

Summarizing, the K-theory group \rref{es4} and \rref{es8} classify all 
stable $D$-brane
charges in Type IIA and TypeIIB on a flat $\IZ_N$ orbifold for $N$ even.

The spectrum of  $D$-branes in flat and toroidal  $\IZ_2$ and $\IZ_4$ 
orbifolds was analyzed
systematically using boundary states and the K-theory formalism  in 
\cite{Gaberdiel:1999ch}. We
briefly  review the results of this construction for  the flat $\IZ_2$ 
orbifold.
To begin with, we  consider the space-time of the form
$$\mathfrak{O}=\IR^{1,5}\times \IR^4/\IZ_2.$$
The generator $g$ of $\IZ_2$ acts as a reflection ${\cal I}_4$ on the 
coordinates of $\IR^4$, and
it acts trivially  or $\IR^{1,5}$. As before $x^0$ and $x^9$ denotes the 
light-cone coordinates. To
describe a D$p$-brane we use the  notation D$(r;s)$, where $p=r+s$ where $r$ 
is the number of Neumann
directions on which $\IZ_2$ acts trivially and $s$ denotes the number of 
Neumann directions
reflected by the group. In the mathematical language, this is a natural 
basis for the decomposition
in terms of the eigenspaces or  action of the representations on the 
coordinates (see section 4.1).

The construction of boundary states follows section \ref{sec:boundarystates} 
with the added feature that
now there will also be  boundary states $\ket{B(r,s)}_{\NN,\Tw}$ and  
$\ket{B(r,s)}_{\RR,\Tw}$\,, constructed from
the NS-NS twisted and R-R twisted sectors, respectively.  To  construct 
consistent $D$-branes,
physical boundary states must be invariant under the combined action of GSO- 
and orbifold-
projection. This restricts the values of $r$ and $s$. The boundary states in 
the different
sector of the theory transforms as

\begin{table}[h]
\centering
\begin{tabular}{|c|c|c|l|} \hline
boundary state &$(-1)^F$ & $(-1)^{\tilde F}$ &${\cal I}_n$ \\ \hline
$\ket{B(r,s)}_{\NN}$    & $+1$              & +1   & +1 \\
$\ket{B(r,s)}_{\RR}$      & $+1$              & $(-1)^{r+s+1}$ & $(-1)^s$ \\
$\ket{B(r,s)}_{\NN,\Tw}$ & $+1$              & $ (-1)^{s}$ & $(-1)^s$ \\
$\ket{B(r,s)}_{\RR,\Tw}$   & $+1$              & $(-1)^{r+1}$ & +1 \\\hline
\end{tabular}
\caption{Action of the GSO and $\IZ_2$ orbifold operators on the boundary
states.}
\label{tab:invariant}
\end{table}
\smallskip
This information allows one to construct the invariant $D$-branes as linear 
combination
of the different invariant boundary states.
The orbifold theory has stable BPS branes ({\it fractional branes}) of the 
form

\bea
\label{eq:bpsz2}
\ket{D(r,s)}&=&\ket{B(r,s)}_{\NN} + \epsilon_1 \ket{B(r,s)}_{\RR}\\\nn
& & + \epsilon_2( \ket{B(r,s)}_{{\NN},{\Tw}} + \epsilon_1 \ket{B(r, 
s)}_{{\RR},{\Tw}})\,,
\eea
defined up to normalization constants deduced by the open/closed string 
duality. This brane carries
two charges due to the coupling of the brane with the fields in the 
untwisted  and twisted R-R sectors. The
sign of the charge is determined by $\epsilon_{1,2}=\pm 1$.  The tree-level 
amplitude of this brane gives
rise to an open string amplitude, as required by Eq.\eqref{eq:treeopen}, 
with
trace $\Tr_{\text{NS-R}}\frac{1+(-1)^F}{2}\frac{1+g}{2}$. The GSO operator 
in
this  partition function projects out the tachyon and makes the spectrum of 
the open
string supersymmetric. Then the brane is stable and  BPS. From Table 
\ref{tab:invariant}  one
can see that this brane exist in Type IIB string theory  for $r$ odd and $s$ 
even and in
Type IIA, $r$ and $s$ should be both even. Note that the corresponding 
K-group is given by \rref{es4}.

The other kind of branes  are non-BPS (called {\it truncated } brane in 
\cite{Gaberdiel:1999ch}). They couple
only to the NS-NS untwisted and R-R twisted sector  and are defined  by
\be
\label{eq:nonbpsz2}
\ket{\hat D(r,s)} =\ket{B(r,s)}_{\NN}
+ \epsilon\;
\ket{B(r,s)}_{\RR,\Tw}
\,.
\ee
This brane carries only one charge represented by $\epsilon$ and  satisfy 
the relation \eqref{eq:treeopen} if
the open string amplitude has the projection operator $\hat 
P=\frac{1+(-1)^Fg}{2}$. This is a slight modification
of the GSO condition since it incorporates  the element $g$. In Type IIB 
this brane is GSO and orbifold invariant
if $r$ and $s$ are both odd and in Type IIA $r$ should be even and $s$ odd. 
The open string spectrum has tachyons
coming from the term $\Tr_{\text{N}}$ and $\Tr_{\text{N}}(-1)^{F}g$ 
respectively and they cancel if and only
if $s=0$. Since the physical conditions restrict $s$ to be odd in both Type 
IIA and Type IIB, the non-BPS brane
is unstable and therefore decays to another brane with the same charge. 
The K-theory group is given in \rref{es8}.

In \cite{Gaberdiel:1999ch} it was shown that non-BPS branes 
\eqref{eq:nonbpsz2} can be stable in $\IZ_2$ orbifolds
with generator $g={\cal I}_4(-1)^{F_L}$ where ${\cal I}_4$ is a reflection 
of the coordinates in $\IR^4$
and  $(-1)^{F_L}$ acts  as $\pm 1$ on the left-moving space-time bosons and 
fermions, respectively. It comes
form the fact that in this kind of orbifolds the consistency conditions put 
both $r$ and $s$ to be even for
non-BPS branes in Type IIB while $r$ should be odd  and $s$ even in Type 
IIA.

\subsection{Branes in $\IZ_N$ orbifolds with $N$ odd}

It is important to note in \ref{subsec:kth} that when $|G|$ is odd, every
non-trivial irreducible real representation has a complex
structure. This means that it is obtained from
a complex representation by forgetting the complex structure.
Accordingly, in the computations below, the allowed branes are
those with all $s_i$ even. So equation \rref{es4}
classifies all orbifold $D$-brane
charge groups in our setting in the case of $|G|$ odd.

We would like to test this prediction analyzing the spectrum of $D$-branes 
in the particular orbifold $\IC^3/\IZ_3$ using boundary states. $D0$-branes 
and their T-duals in this model were analyzed in \cite{Gomis:2000ej} and $D0$ 
and $D3$ branes in the compacified  version in \cite{Hussain:1997hz}. We set 
the construction of these works in a systematic form that allows one to 
classify all branes in this $\IZ_3$ orbifold.

For this model the space-time has the form
$$\mathfrak{O}=\IR^{1,1}\times\IR^2 \times \IC^3/\IZ_3.$$
The light-cone coordinate $x^0$ and $x^9$ define the two dimensional 
Minkowski space. The couple  $(x^1,x^2)$ describe the coordinates on $\IR^2$ 
where $\IZ_3$ acts trivially. On $\IC^3$ we have the complex coordinates
$$
z^i=\frac{1}{\sqrt{2}}(x^{2i+1}+ix^{2i+2})\;  \qquad i=1,2,3.
$$

Non-trivial elements of $\IZ_3$ are denoted by $g^m$ with $m=1,2$ and $g$ 
the generator
acting on the complex coordinates as
$$
g:(z^1,z^2,z^3) \rightarrow (e^{2\pi i \nu_1 }z^1,e^{2\pi i \nu_2} 
z^2,e^{2\pi i \nu_3} z^3)
$$
where $\nu_i=\frac{a_i}{3}$ with $a_i \in \IZ$. To preserve some 
supersymmetry, $a_1+a_2+a_3=0 \;\text{mod}\; 3$. We will be
interested in the case $\nu=(\frac{1}{3},\frac{1}{3}, -\frac{2}{3})$.
The world-sheet fields along $\IC^3$ will be also complexified in the same 
way.
As usual, in the untwisted sector the NS-NS has half-integer modes and the 
R-R
sector the modes are integer. The untwisted R-R ground state is degenerated 
due to the zero-modes along the directions $x^1,x^2,z^1,z^2,z^3$.

In the twisted NS-NS sector fermion variables are modded as $\IZ + \half 
+m\nu_i$. The twisted
R-R sector has modes $\IZ + m\nu_i$ along the directions on which the 
orbifold acts, and there
are zero-modes only on the directions not affected by the orbifold. The 
twisted R-R ground state
is therefore  two-degenerate.

We say that a  $Dp$-brane is of type $(r, {\bf s})$ where  ${\bf 
s}=s_1+s_2+s_3$, it means
that the brane has $r$ number of Neumann boundary conditions along  the 
directions $(x^1,x^2)$
and   $s_1,s_2,s_3$ Neumann directions along  $(x^3,x^4),(x^5,x^6), 
(x^7,x^8)$ directions, respectively. The
case $s_i=1$ corresponds to mixed boundary conditions along any of these 
couple of coordinates.

A detailed analysis of the boundary states and the action of the orbifold 
and the GSO operator
in the  different boundary states of the theory is given in 
\ref{sec:stringappendix}. The action of GSO on
the untwisted sector is the same as in the case discussed in 
\ref{subsec:10d}.  In the twisted
sectors, non-trivial solution to the boundary conditions restrict the values 
of all $s_i$ to be even.  The
conditions for the invariance of  the boundary states have been collected in 
Table \ref{tab:invstatesZ3}.

\begin{table}[h]
\centering
\begin{tabular}{|c|c|c|}
\hline
  & GSO invariant & Orbifold invariant  \\
\hline
$\ket{B(r,{\bf s})}_{\NN}$ & for any $r,{\bf s}$ &  for any   $r,{\bf s}$\\
\hline
$\ket{B(r,{\bf s})}_{\RR}$ & if $r+{\bf s}$  &    \text{ for all $s_i$ 
even}\\
& \text{is even/odd in IIA/IIB} &  \\
\hline
$\ket{B(r,{\bf s})}_{\NN,\Tw}$ & for any  $r$ & for any $r$  \\
\hline
$\ket{B(r,{\bf s})}_{\RR,\Tw}$ & $r$ \text{is even/odd in IIA/IIB} &   for 
any  $r$\\
\hline
\end{tabular}
\caption{Conditions for GSO and orbifold invariant of the different boundary 
states}
\label{tab:invstatesZ3}
\end{table}

We are now  ready to describe the spectrum of $D$-branes. 
The only BPS brane is

\begin{multline}
\ket{D(r,{\bf s})} = \ket{B(r,{\bf s}),\eta}_{\NN} +\epsilon_0 \ket{B(r,{\bf 
s}),\eta}_{\RR}\\
+ \sum_{m=1,2}\epsilon_{m} \left( \ket{B(r,{\bf s}),\eta}_{\NN,\Tw\, g^m} + 
\epsilon _0\ket{B(r,{\bf s}),\eta}_{\RR,\Tw \,g^m}\right)
\end{multline}
where $\epsilon_0$ denotes the sign of the charge with respect to the 
untwisted R-R fields and $\epsilon_{m}$ gives
the sign of the charge of the brane when coupling to the twisted R-R sector.
This brane is stable and supersymmetric as can be seen computing the closed 
string interaction between itself. The
interaction amplitudes in the closed and open string sectors were computed 
in \cite{Diaconescu:1999dt} and we will
not repeat them here. The open string spectrum  is invariant under the 
projector
$\frac{1+(-1)^F}{2}\frac{1+g+g^2}{3}$. Such branes exist for all $s_i$ even 
and $r$ has to be odd in Type IIB or
even in Type IIA. This is the only fundamental brane with R-R charge in this
model.\footnote{ There is a non-BPS brane with the same values of  $(r ,{\bf 
s})$ as that for
the BPS brane. This is unstable. The interaction between the BPS and the 
non-BPS ensures that
the BPS brane is the fundamental brane in the sense that it  is of smaller  
mass and charge
and is stable \cite{Gaberdiel:1999ch}.} This result agrees with the K-theory 
prediction.

\subsection{$D$-brane charges in $\mathbb{Z}_2\times \ldots \times 
\mathbb{Z}_2$}


To give a mathematical discussion for the $(\Z/2)^n$ case,
we refer the reader to the Appendix \ref{sec:mathappendix} for background.
We know by \rref{es17} that since at most $4$ non-trivial
irreducible real representations can occur on $M$, it suffices to
consider $n\leq 4$. Moreover, in the only case of $n=4$ which
cannot be reduced to $n\leq 3$, we have $4$ irreducible real
representations independent in the character group, so
we can use \rref{es18} (with $n=k=4$). Similarly, in the
case of $n=1$ (which was previously treated in the literature),
only one non-trivial real representation exists, so this case
again can be handled by \rref{es18} with $n=1$.
Therefore, we are now
reduced to $n=2$ or $3$.

\vspace{3mm}
The first nontrivial case which cannot be settled by \rref{es17}
or \rref{es18} occurs when $n=2$, and $V$ is the sum
of the three nontrivial $1$-dimensional real representations
$\alpha,\beta, \gamma$ (of course, $\gamma\cong\alpha\otimes\beta$).
Let $A=Ker(\alpha)$, $B=Ker(\beta)$, $C=Ker(\gamma)$, so
$A,B,C$ are the three subgroups of $G$ of order $2$.

To tackle this case, consider the cofibration
\beg{es20}{G/C_+\wedge S^{\alpha+\beta}\r S^{\alpha+\beta}\r S^{
\alpha+\beta+\gamma}.
}
The equivariant $K_G$-theory of the first two summands
can be calculated by \rref{es18} and \rref{es5b}: we have
$\epsilon=0$, and
\beg{es21}{\tilde{K}^{0}_{G}(G/C_+\wedge S^{\alpha+\beta})=
\tilde{K}^{0}_{C}(S^{2\alpha})=\tilde{K}^{0}_{C}(S^0)=\Z\oplus \Z,
}
while
\beg{es22}{\tilde{K}^{0}_{G}(S^{\alpha+\beta})=\Z
}
by \rref{es18}. So we are done if we can calculate the
map from \rref{es22} to \rref{es21} induced by the first
arrow \rref{es20}. In fact, note that the interesting information
is just the image of that map, which can be calculated in
$C\cong\Z/2$-equivariant $K$-theory. When restricted to $C$,
$\alpha\cong\beta=\omega$, which will denote the sign representation
of $C$. Now in $K^{0}_{C}(S^{\omega})$,
we have the element $c$ which, under the inclusion
\beg{es23}{S^0\subset S^{\omega},}
restricts to
\beg{es24a}{1-\omega\in R(C)=\tilde{K}^{0}_{C}(S^0).}
Then the restriction of the generator of $\tilde{K}^{0}_{G}(S^{\alpha+
\beta})$ to $\tilde{K}^{0}_{C}(S^{2\omega})$ is $c^2$.
Our question is thus equivalent to finding the image of
$c^2$ in $\tilde{K}^{0}_{C}(S^0)$ under Bott periodicity.
Now what happens is that the Bott element
$$u\in \tilde{K}^{0}_{C}(S^{2\omega})$$
under the map induced by the inclusion
$$S^{\omega}\subset S^{2\omega}$$
maps to $c$. (This is seen directly by the construction of
the Bott element as $1-H$ where $H$ is the tautological
line bundle on $\C P^1$ where $\Z/2$ acts by minus
on $\C\subset \C P^1$.)
So, we need to find the image of $u^2\in\tilde{K}^{0}_{C}(S^{4\omega})$
under the composition $\beta f^{*}$ where
$$f:S^{2\omega}\r S^{4\omega}$$
is the inclusion (all such inclusions are homotopic) and
$\beta$ is Bott periodicity. But we already know that
$$f^*(u^2)=u(1-\gamma),$$
so
\beg{es24}{\beta f^*(u^2)=1-\gamma.
}
This generates a direct summand in \rref{es21}. In other words,
the first map \rref{es20} induces in $G$-equivariant $K$-theory
the inclusion of a direct $\Z$ summand, and hence for
$V=\alpha+\beta+\gamma$, we have $\epsilon=1$ and
\beg{es25}{\tilde{K}^{1}_{G}(S^{\alpha+\beta+\gamma})=\Z.
}
As already remarked, \rref{es18} implies that for $V\subseteq \alpha
+\beta$, $\epsilon=0$ and
\beg{es26}{\tilde{K}^{0}_{G}(S^{\alpha+\beta})=\Z,\;
\tilde{K}^{0}_{G}(S^{\alpha})=\Z\oplus \Z,\;
\tilde{K}^{0}_{G}(S^0)=\Z^4.
}
All other cases for $n=2$ are related to these by symmetry,
so the case of $n=2$ is completely settled.


In the case of $n=3$, let us denote $\Gamma=(\Z/2)^3$ and
let $\alpha_1,\alpha_2,\alpha_3$ be three $1$-dimensional
real representations of $\Gamma$ which are independent
in the character group. Then similarly to \rref{es26},
again, by \rref{es18}, 
\beg{es27}{\tilde{K}^{0}_{\Gamma}(S^0)=\Z^8,\;
\tilde{K}^{0}_{\Gamma}(S^{\alpha_1})=\Z^4,\;
\tilde{K}^{0}_{\Gamma}(S^{\alpha_1+\alpha_2})=\Z^2,\;
\tilde{K}^{0}_{\Gamma}(S^{\alpha_1+\alpha_2+\alpha_3})=\Z
}
and $\epsilon=0$ in all these cases.  
In the first non-trivial
case, we see by \rref{es17} that
\beg{es28}{\tilde{K}^{1}_{\Gamma}(S^{\alpha_1+\alpha_2+\alpha_3+
\alpha_1\alpha_2})=\Z
}
and $\epsilon=1$ in this case. In the other case
$$V=\alpha_1+\alpha_2+\alpha_3 +\alpha_1\alpha_2\alpha_3,$$
we use the cofibration sequence
\beg{es29}{S^{\alpha_1\alpha_2\alpha_3}\wedge \Gamma/Ker(\alpha_1
\alpha_2\alpha_3)_+\r S^{\alpha_1+\alpha_2+\alpha_3}\r
S^{\alpha_1+\alpha_2+\alpha_3+\alpha_1\alpha_2\alpha_3}.
}
By \rref{es5b}, the $K$-theory of the first term is
$$K^{*}_{Ker(\alpha_1\alpha_2\alpha_3)}(S^{\alpha_1+\alpha_2+\alpha_3})
\cong K^{*}_{G}(S^{\alpha+\beta+\gamma})$$
which, as we have seen, is $\Z$ located in odd dimension.
On the other hand, the $K_\Gamma$-theory of the middle term of \rref{es29}
is calculated by \rref{es27}, giving $\Z$ in even dimension.
We therefore see that for dimensional reasons, the first arrow
of \rref{es29} must induce $0$ in $K_{\Gamma}$, thus giving
\beg{es30}{\tilde{K}^{0}_{\Gamma}(S^{\alpha_1+
\alpha_2+\alpha_3 +\alpha_1\alpha_2\alpha_3})=\Z, \;\epsilon=0.
}
Now all cases of $K_\Gamma$-theories of $1$-point compactifications
of representations of $\Gamma$ which contain at most $4$ different
$1$-dimensional real representations are related to one of the
cases \rref{es27}, \rref{es28} or \rref{es29} by symmetry, so
the case $n=3$ is also completely settled.

The orbifold theories of type  $(\IZ/N)^n$ are very interesting since they 
accept discrete torsion. But so far we have been analyzing orbifolds without 
discrete torsion. In the following we shall revisit the D-brane charge 
classification with the boundary state formalism. The case with discrete 
torsion will be presented in the next section. In the discussion below we will 
concentrate in the  non-compact orbifold $\IZ_2 \times \IZ_2$   without 
discrete torsion. Boundary states and K-theory classifying $D$-brane charges 
in  non-compact and compactified orbifolds $\IZ_2 \times \IZ_2$ were 
analyzed extensively in \cite{Stefanski:2000fp}.

The generators of this group are given by $g_1$ and $g_2$
and they act on the coordinates $x^3, \ldots, x^8$ as

\bea
\label{e50}
g_1(x^0, \ldots, x^9)= (x^0, x^1,x^2,x^3,x^4,-x^5,-x^6,-x^7,-x^8,x^9)\\\nn
g_2(x^0, \ldots, x^9)= (x^0, x^1,x^2,-x^3,-x^4,x^5,x^6,-x^7,-x^8,x^9)
\eea
with $g_3=g_1 g_2$.
To describe a D$p$-brane we use the notation $(r;{\bf s})=(r;s_1,s_2,s_3)$, 
where $p=r+s_1+s_2+s_3$. This means
that the brane has Neumann boundary conditions along $r+1,s_1,s_2,s_3$  of 
the  $(x^0,x^1,x^2,x^9)$, $(x^3,x^4)$,
$(x^5,x^6),(x^7,x^8)$, respectively.
To connect with our mathematical notation above, the coordinates
$x^0,x^1,x^2$ and $x^9$ are copies of the representation $1$,
the coordinates $x^3, x^4$
are copies of $\alpha$, the coordinates $x^5,x^6$ are copies
of $\beta$ and $x^7, x^8$ are copies of $\gamma$ (although
$\alpha$, $\beta$, $\gamma$ are notationally interchangeable).

There are several kinds of branes. We will write down only the fundamental 
branes
that carry the smaller charge and mass. It was noticed in 
\cite{Gaberdiel:2000fe} that
in the  kind of orbifold in question, the open string endpoints can carry 
the  conventional   and  projective
representations of the orbifold group; and that the presence of discrete 
torsion does not change the results\footnote{We recall that the conventional 
representation corresponds to the case when $\gamma(g)\gamma(h)=\gamma(gh))$.}. The 
boundary states carrying the 
projective representation were analyzed in \cite{Craps:2001xw,Gaberdiel:2000fe}.

In all branes given below  $r$ is even or odd in Type IIA or Type IIB, 
respectively. There are two
types of BPS branes. One is a {\it fractional} brane with projective 
representation of the orbifold group on the Chan-Paton factors. It is 
defined  for all $s_i=1$ and has the  form

\be
\ket{D(r;{\bf s});{\bf a}}=\ket{B(r;{\bf s});{\bf a}} +  \ket{B(r;{\bf 
s});{-\bf a}}\;,
\ee
where

\begin{align}
\ket{B(r;{\bf s});{\bf a}}&=\ket{B(r;{\bf s});{\bf a}}_{\NN;U}+ 
\epsilon\ket{B(r;{\bf s});{\bf a}}_{\RR;U} \\ &+ \epsilon'(\ket{B(r;{\bf 
s});{\bf a}}_{\NN;T_{g_i}}+\epsilon \ket{B(r;{\bf s});{\bf 
a}}_{\RR,T_{g_i}})\nn
\;.
\end{align}
We have to stress that the moduli space of this brane consist of the 
different fixed planes of $g_i$ with $i=1,2,3$. The position of the brane 
along the directions on which the orbifold acts trivially have been dropped 
out and  ${\bf} a$ is the position of the brane in the directions  on the 
 planes fixed by $g_i$. The  orbifold acts on the Chan-Paton factors by a  
projective representation.  This brane carries charge only with respect to 
the untwisted R-R charge. Therefore the respective K-group is given by 
\rref{es25}.

Next, one has a  {\it fractional} brane with conventional representation. It 
is defined  for all $s_i$ even and is given
by
\be
\ket{D(r,{\bf s})}=\ket{B(r,{\bf s})}_{\NN} + \epsilon \ket{B(r,s)}_{\RR}
+\sum_{i=1}^3 \epsilon_i (\ket{B(r,{\bf s})}_{{\NN},{\Tw}_{g_i}} + 
\epsilon_1 \ket{B(r,{\bf s})}_{{\RR},{\Tw}_{g_i}})
\ee
where $\epsilon=\pm 1$ determines the sign of the charge with respect to the 
untwisted
R-R sector, $\epsilon_i=\pm 1, i=1,2,3$ denote the sign of the charge with 
respect to the
R-R twisted by $g_i$. Such brane is stuck at {\it all} fixed point 
$x^3=\ldots=x^8=0$ of $\IZ_2 \times \IZ_2$. This brane  carries four 
charges, $\epsilon$ and $\epsilon_i, i=1,2,3$ and its K-theory group is 
given by the last relation in \eqref{es26}.

For the  non-BPS case, there are also two kind of branes, one is also a 
fractional brane with conventional representation
\be
\ket{\hat D(r,{\bf s})}= \ket{B(r,{\bf s})}_{\NN}  + \epsilon 
\ket{B(r,{\bf s})}_{{\RR},{\Tw}_{g_i}},
\ee
%
%
%
This $D$-brane is charged under a massless R-R field in the twisted sector by 
$g_i$. The corresponding K-theory is determined by the first relation in 
\eqref{es26}.

The last brane is a {\it fractional brane} with conventional representation 
of the group on the endpoints of the open string. It has one $s_i$ odd and 
the rest even. It couples to two of the three R-R twisted sectors. Say for 
instance
$g_i$, and $g_j$ for $i \neq j$. The respective boundary state is
\be
\ket{\hat D(r,{\bf s})}=\ket{B(r,{\bf s})}_{\NN} + 
\epsilon_i\ket{B(r,s)}_{\RR,{\Tw_{g_i}}}
+\epsilon_j \ket{B(r,{\bf s})}_{\RR,{\Tw_{g_j}}}+ \epsilon_i\epsilon_j 
\ket{B(r,{\bf s})}_{{\NN},{\Tw}_{g_k}}
\ee
This brane is stuck at all fixed points of the orbifold group. The K-theory 
group is given by the second relation in \eqref{es26}.

\subsection{Branes in Type II on $T^6/\IZ_2 \times \IZ_2$}

So far  we have analyzed the $D$-brane charge spectrum of the 
non-compactified
$\IZ_2 \times \IZ_2$ orbifold. The compactified case can be obtained 
straightforwardly. From the point of view of $K$-theory, the equivariant
$K$-theory groups relevant here are determined by the
cases discussed above: Suppose we have a transverse
torus $T^{a;b,c,d}$ where the numbers $a,b,c,d$ are
as above. Then we have
$$K^{*}_{\Z/2\times\Z/2,c}(T^{a;b,c,d})=
\bigoplus
\left(\begin{array}{l}a\\a^{\prime}\end{array}\right)
\left(\begin{array}{l}b\\b^{\prime}\end{array}\right)
\left(\begin{array}{l}c\\c^{\prime}\end{array}\right)
\left(\begin{array}{l}d\\d^{\prime}\end{array}\right)
K^{*}_{\Z/2\times \Z/2,c}(\R^{a^{\prime};
b^{\prime},c^{\prime},d^{\prime}}).$$
where the sum is over all $0\leq a^{\prime}\leq a$,
$0\leq b^{\prime}\leq b$, $0\leq c^{\prime}\leq c$,
$0\leq d^{\prime}\leq d$.
The corresponding boundary states are given  as those above in the 
uncompactified case. However in this case, one has to put particular 
attention
to the stability radius.

\section{K-theory and  discrete torsion}\label{sec:discretetorsion}

Mathematically, discrete torsion in not an intrinsic property of the
orbifold but its $K$-theory, which becomes twisted $K$-theory.
This means that $K$-theory varies as we move around the orbifold, i.e.
we have a ``bundle of $K$-theories'' on the orbifold. From a spacetime
point of view, an $H_3$-flux can cause a twisting of $K$-theory in
this sense. From the world-sheet point of view, on the other hand,
$K$-theory twisting corresponds to an automorphism of the category
of vector spaces in which Chan-Paton bundles take place.

It is proved in \cite{as} that
on a $G$-space $X$, $G$-equivariant $K$-theory $H_3$-twistings are 
classified
by elements of
\beg{es31}{H^{3}_{Borel}(X,\Z).
}
Borel cohomology of a $G$-space $X$ is obtained by taking
a space $EG$ which is contractible but has a free $G$-action
(such space is unique up to homotopy equivalence under some
minimal topological assumptions which we do not discuss here).
Then Borel cohomology simply means cohomology of the
Borel construction
$$EG\times X/(y,x)\sim(gy,gx) \;\text{for $x\in X$, $y\in EG$, $g\in G$.}
$$
For our purposes, we must answer the question as to what kind of
twistings are possible in equivariant $K_G$-theory with compact
supports of a space $X$. However, the answer turns out to be
the same, there are twistings corresponding to all elements
of \rref{es31}, in other words no compact supports are needed
in the twisting group. The reason is that generalized cohomology
with compact supports is a direct limit of relative cohomology
groups of pairs $(X,U)$ where $X-U$ is compact. All those
groups are consistently twisted by a given element of
\rref{es31}, hence so is the direct limit.

\vspace{3mm}
In our case, the space $X$ is a representation, hence is
contractible without compact supports. In this case, the
Borel cohomology group \rref{es31} is simply the cohomology
of the group
\beg{es32}{H^3(G,\Z).}

\subsection{$\IZ_2 \times \IZ_2$ orbifold}
We shall perform our mathematical calculation in only one case, namely
$$G=(\Z/2)^2.$$
In this case, we may write
\beg{es32a}{H^*(G,\Z/2)=\Z/2[x,y],\; dim(x)=dim(y)=1.}
Recall that the Bockstein homomorphism
\beg{es32b}{b:H^i(G,\Z/2)\r H^{i+1}(G,\Z)}
is the connecting homomorphism of the long exact sequence
associated with the short exact sequence on coefficients
$$0\r \Z\r\Z\r\Z/2\r 0.$$
(Note: the Bockstein is usually denoted by $\beta$, but
that would conflict with some of our other notation.)

\vspace{3mm}
Now one has
\beg{es33}{H^3(G,\Z)\cong \Z/2
}
where the non-trivial element is
\beg{es33a}{b(xy),}
using the notation of \rref{es32a}, \rref{es32b}.
For the twisting $\tau$ associated with this
element, the relevant twisted $K$-groups are
\beg{es36}{{
\begin{array}{ll}
K^{0,c}_{G,\tau}(*)=\Z, & \text{($\epsilon$ even)}\\
K^{1,c}_{G,\tau}(\alpha)=\Z & \text{($\epsilon$ odd)}\\
K^{1,c}_{G,\tau}(\alpha+\beta)=\Z\oplus \Z & \text{($\epsilon$ odd)}\\
K^{1,c}_{G,\tau}(\alpha+\beta+\gamma)=\Z^4 & \text{($\epsilon$ odd).}
\end{array}
}}

To prove \rref{es36}, the first group is $K$-theory of projective
representations of $G$ with the cocycle given by \rref{es33a}.
Its elements can be thought of, for example, as
virtual representations of
the quaternionic group $\{\pm 1, \pm i, \pm j, \pm k\}$ where
$-1$ acts by $-1$; these are just sums of the quaternionic
representation. Projective equivariant bundles on $S^1$ are
trivial, hence the $K^1$-group vanishes in this case. 
For the three remaining groups, we smash again, in order, with the
familiar cofibration sequences
$$G/A_+ \r S^0 \r S^\alpha,$$
$$G/B_+\r S^0\r S^\beta,$$
$$G/C_+\r S^0\r S^{\gamma}.$$
The key point is that twisting disappears on proper subgroups,
so we get long exact sequences
\beg{esx1}{\r K^{i,c}_{G,\tau}(\alpha)\r K^{i,c}_{G,\tau}(*)\r 
K^{i,c}_{G/A}(*)\r
}
\beg{esx2}{\r K^{i,c}_{G,\tau}(\alpha+\beta)
\r K^{i,c}_{G,\tau}(\alpha)\r K^{i,c}_{G/B}(\alpha)\r
}
\beg{esx3}{\r K^{i,c}_{G,\tau}(\alpha+\beta+\gamma)
\r K^{i,c}_{G,\tau}(\alpha+\beta)\r K^{i,c}_{G/C}(*)\r.
}
The last term in \rref{esx3} is by Bott periodicity.
If we assume that the calculation of $K^{i,c}_{G,\tau}(\alpha)$
is correct, then in both \rref{esx2} and \rref{esx3} the last two
terms shown are known and occur in different dimensions, so
the remaining calculations follow.

\vspace{3mm}
In \rref{esx1}, the last two terms shown are $\Z$, $\Z\oplus\Z$,
and occur both in even dimensions, so it remains to show
that the map between them is the inclusion of a direct summand.
There are several ways of showing this. One is to map into
the respective $K$-groups which we get when we smash all spaces
involved with $EG_+$. Both groups inject, and in the target,
the groups reduce to non-equivariant $K$-groups:
\beg{esx4}{K^{i}_{\tau}(BG)\r K^{i}(B(G/A)).}
Both groups are known, in fact \rref{esx4} has the form
\beg{esx5}{\Z\r \Z\oplus \Z_2}
($\Z_2$ means $2$-adic numbers). There are Atiyah-Hirzebruch
spectral sequences converging to both groups. In the case
of $K^{*}(B(G/A))$, the AHSS collapses, but the
generator $\iota$ of $H^0
(B(G/A),\Z)$ supports an extension to higher filtration
degrees. Thus, $2\iota$ generates a direct summand $\Z$. In the twisted 
case,
the Atiyah-Hirzebruch spectral sequence is described in \cite{as},
and the first differential is calculated as
$$d_3=b Sq^2 + H_3$$
(recall that we denoted the Bockstein by $b$ to avoid conflict with
other notation).
In the case
of $K^{*}_{\tau}(BG)$, $H_3$ is multiplication by \rref{es33a}, so
the generator $\iota$ of $H^0$ actually supports a differential.
However, the odd degree elements of the $E_2$
Atiyah-Hirzebruch spectral sequence (which are all copies of $\Z/2$)
are then entirely wiped out
by $d_3$, so no further differentials are possible, showing that
$2\iota$ is a permanent cycle which must
generate $K^{*}_{\tau}(BG)$ because it is in filtration degree $0$.
It maps to $2\iota$ in \rref{esx4}, thus showing that \rref{esx4}
is an inclusion of a direct summand $\Z$, and hence so is
the second map \rref{esx1}.

\vspace{3mm}
There is another, more conceptual reason why \rref{es36} holds.
The reader may notice that the twisted groups \rref{es36},
in the homotopy-theoretical language, appear
shifted from the corresponding groups \rref{es25}, \rref{es26}
by the dimension
\beg{es400}{1+\alpha+\beta+\alpha\beta.}
This is indeed the case and is part of a general pattern. For
any space $X$ and any real even-dimensional vector bundle $\eta$
on $X$, we have the induced non-equivariant bundle on the
Borel construction, whose Stiefel-Whitney classes can be thought
of as the equivariant Stiefel-Whitney classes of $\eta$ in
Borel cohomology. In particular, there is the class
$$W_3(\eta)=bw_2(\eta)\in H^{3}_{Borel}(X,\Z)$$
which is the obstruction to $\eta$ being $Spin^c$. But
in addition to that, if we denote by $V$ the total space of $\eta$,
we have generalized Bott periodicity in twisted $K$-theory
\beg{es401}{K^{i,c}_{G,\tau}(V)\cong K^{i,c}_{G,\tau+W_3(\eta)}(X).}
The proof just mimics the classical index-theoretical proof
of Bott periodicity in the world of twisted bundles
(see also \cite{donk}).

\vspace{3mm}
In the current setting, $X=*$ and $V$ is given by \rref{es400},
and one easily sees that $W_3(V)$ is equal to \rref{es33a};
recall that the total Stiefel-Whitney class is
$$w(V)=(1+x)(1+y)
(1+x+y)).$$
Thus, \rref{es401} implies the shift indicated.
It is worth remarking that all twistings of $(\Z/2)^n$-equivariant
$K$-theory over a point are of this form, and
hence this method can be used to find all twisted $K_{(\Z/2)^n}$-theory
groups with compact support of representations. Details will
be given elsewhere. 
The boundary states for this case are those described in the case  of orbifold without 
discrete torsion but the role of the $s_i$ is exchanged.  The K-theoy 
corresponding to these branes is given \rref{es36}.

In \cite{Gaberdiel:2000jr,Craps:2001xw} it was noticed that there is a ``T-duality" relating the
theory without and with discrete torsion. On the D-branes, this duality leaves $r$ invariant while 
if  the $s_i$ are even, they becomes odd and vice versa.  From the K-theory point of view, the 
K-group  \rref{es36} is suggestive because of ``T-duality'' with the picture
\rref{es25}, \rref{es26}. $K$-theoretically, this T-duality results by adding the
representation $1+\alpha+\beta+\gamma$ (as elsewhere, we denote
a trivial representation by the same symbol as its dimension,
so $1$ is the $1$-dimensional real representation),
which is orientable, but its $W_3$ is the non-trivial
class in $H^3(\Z/2\times \Z/2,\Z)$. In particular, the shift $1+\alpha+\beta+\gamma$ gives 
an isomorphism between  K-group $\tilde K^0_G(S^0)$ and $\tilde K^{1,c}_{G,\tau}(\alpha+\beta+\gamma)$. 
The conventional representations of the brane classified by $\tilde K^0_G(S^0)$  is preserved under this shift.
The reason is that suspension by one of the representations
$\alpha, \beta, \gamma, \alpha+\beta$, or $\alpha + \beta +
\gamma$ in this case corresponds roughly to replacing the group
$\IZ_2 \times \IZ_2$ by the subgroup. But on subgroups, the central extension
given by this cocycle becomes trivial (the cocycle becomes a
coboundary), which is why the representation becomes conventional. 
A similar discussion follows for the BPS brane with projective representation.

\subsection{$\IZ_N \times \IZ_N$ orbifold with $N$ odd}

To check the $K$-theory prediction here, let us note
again that the group of discrete torsions is $\Z/N$.
We discuss the case when the torsion is given by the
generator of that group. In this case, the
relevant twisted $K$-group is $\Z$. To see this,
we can use the twisted Atiyah-Hirzebruch spectral sequence converging
to the corresponding completed $K$-group. The $E_2$-term is
\beg{ech1}{\Z[x,y]/(Nx,Ny)\oplus\Z/N[x,y]\omega,\; dim(x)=dim(y)=2,
dim(\omega)=3.
}
By \cite{as}, there is a $d_3$-differential
$$d_3:1\mapsto \omega,$$
so
$$d_3(x^m y^n)=\omega x^m y^n,$$
and the $E_3$-term is $\Z$, generated by $N\cdot 1$. The numbers
$s_1,s_2,s_3$ are even here.

Our results above agree with those found in \cite{Craps:2001xw} where the 
boundary states for this kind of orbifold were constructed. The brane 
corresponding to the K-theory group is a fractional brane with projective 
representation where all $s_i$ are even. It has the form

$$
\ket{D(r,s_1,s_2,s_3);{\bf a},\epsilon, \epsilon_1} = 
\sum_{m=0}^{N-1}\epsilon^m_1
\sum_{n=0}^{N-1}w^{-mn}\ket{D(r;s_1,s_2,s_3);, g^n_2 {\bf a},\epsilon}
_{Tg_1^m}
$$
This brane carries a charge under the untwisted R-R charge. But
in addition to this brane, there is a {\it bulk} brane with the same form as 
the projective fractional brane but with all $s_i$ equal to 1. Because of 
the different boundary conditions for each couple of coordinates, it is hard 
to predict the action of the orbifold on the Chan-Paton factors. However 
this bulk brane carries charge under the untwisted R-R field, too.

\subsection{$\IZ_N \times \IZ_N$ orbifold with $N$ even}

In this case, too, the $K$-theory prediction can be
calculated. However, in some sense, the case considered
here is the opposite to the case of $\Z/N\times \Z/N$ with
$N$ odd, since in the present case we do not consider minimal
torsion, but torsion corresponding to the subgroup
$\Z/2\subset \Z/N$. In this case, the torsion has the
form of a $W_3$-class of a bundle, so we
can use the Thom isomorphism of Donovan-Karoubi
(see above and \cite{donk}): The predicted group of
charges is then
$$\Z^{N^{2}/2^m}$$
where $0\leq m\leq 2$. The number $m$ is the same
as the corresponding number for the untwisted group
when we add $1$ to each of the numbers $r,s_1,s_2,s_3$.

\section{Conclusions}\label{sec:conclusions}

Using perturbative string theory we have discussed 
the construction of branes in various orbifolds including with 
discrete torsion. We have shown full agreement with the corresponding equivariant and twisted K-theory results. 
Although a large part of our discussion is present in the literature we have aimed at presenting 
a comprehensive picture of both the physical and the mathematical sides. We have also obtained new 
results, most prominently the full analysis of the $\mathbb{Z}_3$ orbifold and the K-theory of the 
corresponding models of $\mathbb{Z}_N\times \mathbb{Z}_N$ orbifolds without and with discrete torsion. 
We have also presented a full K-theoretic treatment of $(\mathbb{Z}_2)^n$ orbifolds with $n\le 4$.

The asymmetry for $N$ even or odd predicted by K-theory presents interesting implications for the 
spectrum of $D$-brane charges and deserves further study. For example, in the study of discrete symmetries 
in quiver gauge theories dual to $D3$-branes on orbifolds \cite{heisenberg} and asymmetry for $\mathbb{Z}_N$ 
orbifolds was also found depending on whether $N$ is odd or even; this asymmetry is arguably related to the 
structures of section \ref{sec:orbifoldsexamples}, in particular, to the homomorphism of subsection 
\ref{subsec:kthNeven}.

In the context of perturbative string theory a lot of progress can be made in the classification of 
$D$-brane charges because one can directly compute them. However, there are many situations where a 
perturbative description is not available. There are many interesting questions that arise in this context,  
here we list some of those we hope to address in the near future.  

The AdS/CFT correspondence has proved useful in attacking some
questions related to $D$-brane charges \cite{heisenberg,Gukov:1998kn} since it provides a 
dual description of
string theory in the presence of Ramond-Ramond fluxes. We plan to extend 
part of our discussion to that
situation. A particularly interesting situation, raised in \cite{Burrington:2007mj}, is the 
structure of string theory in nonabelian orbifolds and its relation to the decoupling limit and the 
quiver gauge theory.

\section*{Acknowledgments}

We would like to thank D. Belov, M. Gaberdiel, E. Gimon, H. Klemm and B. Uribe for 
comments, suggestions and correspondence.
N.Q. wants to thank UC Berkeley, LBNL and  MCTP for
hospitality during the early stages of this project. N.Q was supported by UC 
Mexus and Conicet. This work
is  partially supported by Department of Energy under grant 
DE-FG02-95ER40899 to the University of Michigan.


\appendix

\section{Perturbative string theory}\label{sec:stringappendix}
In this section we collect a number of relations that we used in the 
derivation of the main results of the text.

\subsection{Hamiltonians}

Along these notes, we will work with $\alpha'=1$.
The open string Hamiltonian is given as
\be
H_0 = \pi {\bf p}^2 +\frac{1}{4\pi}{\bf w}^2+ \pi \sum_{\mu=1}^9 \left( 
\sum_{n=1}^{\infty} \alpha^{\mu}_{-n}\alpha^{\mu}_n + \sum_{r>0}r 
\psi^{\mu}_{-r}\psi^{\mu}_r\right) + \pi C_0,
\ee
where ${\bf p}$ is the momentum of the endpoint open strings along the 
Neumann directions and ${\bf w}$ denotes the difference between the two 
endpoints. The zero-point energy $C_0$  is zero in the R sector while in  
the NS-sector  it is equal to $-\half +\frac{s}{8}$ with $s$ the number of 
Dirichlet-Neumann boundary conditions.

The closed string Hamiltonian is

$$
H_c = \pi {\bf K}^2 + 2 \pi \sum_{\mu = 1}^8 \left( \sum_{n=1}^{\infty}
(\alpha_{-n}^{\mu}\alpha^{\mu}_n +{\wt \alpha}_{-n}^{\mu}{\wt 
\alpha}^{\mu}_n )
+ \sum_{r>0}r(\psi^{\mu}_{-r}\psi^{\mu}_r + {\wt \psi}^{\mu}_{-r} {\wt 
\psi}^{\mu}_r ) + 2\pi C_c
\right)
$$
where $C_c$ is equal to -1 in the NS-NS sector and 0 in the R-R sector; and ${\bf K}$ is the closed 
string momentum. 

\subsection{ The R-R ground states}
\label{app:rrgs}

The boundary conditions defining the R-R vacuum are given by  
(\ref{eq:fermionbc}) for $r=0$

\begin{align}
\label{eq:zmbc}
(\psi^{\mu}_0+i\eta \wt \psi^{\mu}_0)\ket{Bp,\eta}^0_{\RR}=0\,,\quad \quad 
\mu = 1, \ldots, p+1\nn\\
(\psi^{\nu}_0-i\eta \wt \psi^{\nu}_0)\ket{Bp,\eta}^0_{\RR}=0\,,\quad \quad 
\nu = p+2, \ldots, 8\;.
\end{align}
Let us to introduce
\be
\psi^{\mu}_{\pm}= \frac{1}{\sqrt{2}}(\psi^{\mu}_0 \pm i\wt \psi^{\mu}_0)\;,
\ee
in these variables, equations \eqref{eq:zmbc} define the R-R ground 
state
$\ket{Bp,k,\eta}^0$ by the conditions
\begin{align}
\label{eq:zmpm}
\psi^{\mu}_{\eta}\ket{Bp,k,\eta}^0_{\RR}&=0  \quad \mu=1,\ldots, p+1\nn\\
\psi^{\nu}_{-\eta}\ket{Bp,k,\eta}^0_{\RR}&=0  \quad \nu=p+2,\ldots, 8
\end{align}
where $\eta=\pm $ and  $\psi^{\mu}_{\eta}$ and $\psi^{\nu}_{-\eta}$ can be 
seen as annihilation  operators in the Neumann and Dirichlet directions 
respectively. The state $\ket{Bp,k,-\eta}$ is created by applying 
consecutively creation operators on the R-R ground state
\be
\ket{Bp,k,-\eta}^0_{\RR}=\prod_{\mu=1}^{p+1}\psi^{\mu}_{-\eta}\prod_{\nu=p+2}^{8}\psi^{\nu}_{\eta}\ket{Bp,k,\eta}.
\ee

The representation of the GSO operator in the R-R zero modes is given as 
\cite{Bergman:1997rf}

\begin{align}
(-1)^F &= \prod_{\mu=1}^9 (\sqrt{2}\psi^{\mu}_0) = \prod_{\mu=1}^9 
(\psi^{\mu}_+ + \psi^{\mu}_-),\nn\\
(-1)^{\wt F} &= \prod_{\mu=1}^9 (\sqrt{2}\wt \psi^{\mu}_0) = \prod_{\mu=1}^9 
(\psi^{\mu}_+ - \psi^{\mu}_-).\nn\\
\end{align}
The action of these operators on the R-R ground state is
\begin{align}
\label{eq:gsogs}
(-1)^{F}\ket{Bp,k,\eta}_{\RR}^0&= \ket{Bp,k -\eta}_{\RR}^0\, ,\nn\\
(-1)^{\wt F}\ket{Bp,k,\eta}_{\RR}^0&= (-1)^{p+1} \ket{Bp,k -\eta}_{\RR}^0\,.
\end{align}

\subsection{ Boundary states in $\IZ_3$ orbifold}

\subsubsection{The untwisted sector}
\label{app:uts}

The world-sheet fields along $\IC^3$ are defined as

\begin{align}
Z^i&=\frac{1}{\sqrt{2}}(X^{2i+1} +iX^{2i+2}) &{\bar 
Z}^i&=\frac{1}{\sqrt{2}}(X^{2i+1} - iX^{2i+2}) \quad i=1,2,3 \nn \\
\lambda^i&=\frac{1}{\sqrt{2}}(\psi^{2i+1} +i\psi^{2i+2})  &
{\bar \lambda}^i&=\frac{1}{\sqrt{2}}(\psi^{2i+1} - i\psi^{2i+2}) \;.
\end{align}
The mode expansion of these complex fields are
\begin{align}
Z^i & = z^i + 2\pi p^i \tau + \frac{i}{\sqrt{2}}
\sum_{n\neq 0}\frac{1}{n}\left(\beta^i_n e^{-2\pi i n(\tau+\sigma)} + \wt 
\beta^i_n e^{-2\pi i n (\tau + \sigma)}  \right)\;,\nn\\
\lambda^i &= \sqrt{2\pi}\sum_{r}\lambda_r^i e^{-2\pi i r(\tau- \sigma) }\;,
\end{align}
with similar relations for the conjugates. The complex oscillator modes are 
defined in terms of the real modes as
\begin{align}
\beta^i& =\frac{1}{\sqrt{2}}(\alpha^{2i+1}+i\alpha^{2i+2}) & \bar 
\beta^i&=\frac{1}{\sqrt{2}}(\alpha^{2i+1}-i\alpha^{2i+2}) \nn\\
\lambda^i& =\frac{1}{\sqrt{2}}(\psi^{2i+1}+i\psi^{2i+2}) & \bar 
\lambda^i&=\frac{1}{\sqrt{2}}(\psi^{2i+1}-i\psi^{2i+2})
\end{align}
with similar relations for the left-modes. Under quantization the  
(anti)-commutation relations are

\begin{align}
[\beta^i_n,\bar \beta^j_m]&=[\wt \beta^i_n,\wt{\bar 
\beta}^j_m]=n\delta_{n+m}\delta^{ij},\\
\{\lambda^i_r, \bar \lambda^j_s\}&= \{\wt {\lambda}^i_r,\wt{ \bar 
\lambda}^j_s\}=\delta_{r+s}\delta^{ij}.
\end{align}

The generator of the orbifold $\IZ_3$ acts on world-sheet fields as
\begin{align}
\label{eq:wst} 
g: & Z^i \rightarrow e^{2\pi i v_i}Z^i\nn\\
g: & \bar Z^i \rightarrow e^{-2\pi i v_i} \bar Z^i
\end{align}

Under the action of the $\IZ_3$ generator they are transform as
\begin{align}
\label{eq:osct}
\beta^i&  \rightarrow e^{2\pi i \nu_i}\beta^i & \bar \beta^i&  \rightarrow 
e^{-2\pi i \nu_i}\bar \beta^i \nn\\
\lambda^i&  \rightarrow e^{2\pi i \nu_i}\lambda^i & \bar \lambda^i&  
\rightarrow e^{-2\pi i \nu_i}\bar\lambda^i
\end{align}

The boundary conditions for the boundary states can be written explicitly given the 
the boundary conditions it satisfies along each pair of coordinates  
$(x^{2i+1,x^{2i+2}})$ defining the plane $z^i$. If a couple of real 
coordinates has mixed boundary conditions (corresponding to any $s_i=1$) the 
gluing conditions are
\begin{align}
\label{eq:complexND}
(\beta^i_n \pm\wt {\bar \beta}^i_{-n})\ket{\eta}&=0 \nn\\
(\bar \beta^i_n \pm \wt{\beta}^i_{-n})\ket{\eta}&=0  \nn\\
(\lambda^i_r \pm i\eta {\wt{ \bar \lambda}}^i_{-r})\ket{\eta}&=0 \nn\\
(\bar \lambda^i_r \pm i\eta \wt{\lambda}^i_{-r})\ket{\eta}&=0
\end{align}
where the positive sign corresponds to Neumann-Dirichlet boundary conditions 
  and negative sign for Dirichlet-Neumann conditions. The boundary state 
solving these equations  is
\be
\label{eq:bsnd}
\ket{ \eta}_{\substack{\text{N-D}\\
                                       \text{D-N}}}
=\text{exp}\left\{ \mp\sum_{n=1}^{\infty}\frac{1}{n}
(\beta^i_{-n} \wt {\beta}^{i}_{-n}+
\bar \beta^i_{-n} \wt {\bar \beta}^{i}_{-n})
\mp i\eta\sum_{r>0}^{\infty}(\lambda^i_{-r} \wt { \lambda}^{i}_{-r}+
\bar \lambda^i_{-r} \wt {\bar \lambda}^{i}_{-r}) \right\}
\ket{ \eta}^0_{\substack{\text{NS-NS}\\
\text{R-R}}}
\ee
where now the negative  and positive signs in the exponential are associated 
to Neumann-Dirichlet and  vice versa, respectively.

On the other hand, if the couple has the same boundary conditions (when 
$s_i=0,2 $ corresponding to Dirichlete-Dirichlet or Neumann-Neumann, 
respectively) the boundary equations  are
\begin{align}
\label{eq:complexNN}
(\beta^i_n \pm {\wt\beta}^i_{-n})\ket{\eta}&=0\nn\\
(\bar \beta^i_n \pm \wt{\bar \beta}^i_{-n})\ket{\eta}&=0\nn\\
(\lambda^i_r \pm i \eta  {\wt\lambda}^i_{-r})\ket{\eta}&=0\nn\\
(\bar \lambda^i_r \pm i\eta  \wt{\bar \lambda}^i_{-r})\ket{\eta)}&=0\;,
\end{align}
with positive sign for Neumann-Neumann and negative sign for 
Dirichlet-Dirichlet. The boundary state solving the  equations 
\eqref{eq:complexNN} for each par of coordinates is
\be
\label{eq:bsnn}
\ket{ \eta}_{\substack{\text{N-N}\\
                                       \text{D-D}}}=
\text{exp}\left\{\mp
\sum_{n=1}^{\infty}\frac{1}{n}(
\beta^i_{-n}  \wt {\bar \beta}^{i}_{-n}+
\bar \beta^i_{-n} {\wt \beta}^{i}_{-n})
\mp i\eta\sum_{r>0}^{\infty}(
\lambda^i_{-r} \wt{\bar \lambda}^{i}_{-r}+
\bar \lambda^i_{-r}  \wt { \lambda}^{i}_{-r})
\right\}\ket{\eta}^0_{\substack{\text{NS-NS}\\
\text{R-R}}}
\ee
negative and positive signs are associated to Neumann-Neumann and 
Dirichlet-Dirichlet directions, respectively. Note that we 
have dropped out the dependence of $r$, $s$ and the momentum but it will be restored latter. The R-R Fock 
vacuum is defined solving equations \eqref{eq:complexND} and 
\eqref{eq:complexNN} for the zero-modes.

Let us see now if these boundary states are GSO and orbifold invariant.
For the untwisted NS-NS the action of the GSO operator is the same as that 
presented in \ref{subsec:10d}.
The orbifold action on the NS-NS vacuum is trivial. The boundary state 
defined in \eqref{eq:bsnn} has the combination  $\beta^i \wt{\bar 
\beta}^i + \bar \beta^i \wt{ \beta}^i$ (also for fermions) in the 
exponential. It is invariant under the action of the orbifold since each 
oscillator mode is multiplied by a  complex one. However the term
$\beta^i \tilde\beta^i + \bar \beta^i \wt{\bar \beta^i}$ in 
\eqref{eq:bsnd} in not orbifold invariant. Only the combination 
$\frac{1}{3}(1+g\ket{\eta}+g^2\ket{\eta})$ will be invariant in this case.

In the untwisted R-R sector, the GSO has a complex representation on the 
zero-modes. The analysis discussed in \ref{app:rrgs} is straight forward in 
this case and we will not repeat it again. The representation of the  
orbifold on the zero-modes is determined by the cubic roots of unity. 
However, only one of the three   possibilities will leave invariant R-R 
ground states. The relevant representation is
\be
\label{eq:z3zm}
g =\prod_{i=1,2,3}(e^{2\pi i/3}-i\sqrt{3}\,\lambda^{i}_0 \bar 
\lambda^{i}_0)(e^{2\pi i/3}-i\sqrt{3}\,\wt\lambda^{i}_0 \bar 
{\wt\lambda}^{i}_0)\;
\ee
with $g^2$ being the complex conjugate of $g$. It  can be verified easily 
using \eqref{eq:complexND},  \eqref{eq:complexNN} and \eqref{eq:z3zm} that 
only the R-R vacuum with the same boundary conditions is orbifold invariant. 
Then only the boundary state defined in equation \eqref{eq:bsnn} will be 
preserved by the orbifold projection.
\subsubsection{Twisted sector}
In the twisted sector the modes are shifted by $m \nu_i$. The annihilation   
operators are $\beta_{n+m\nu_i}$ and $\bar {\beta}_{n-m\nu_i}$ for the left 
modes and  $\wt{\beta}_{n-m\nu_i}$ and  ${\wt {\bar {\beta}}}_{n+m\nu_i}$ 
for the right modes. The respective  fermionic operators  are defined in a 
similar way. The boundary conditions in the sector twisted by $g^m$ are 
those as \eqref{eq:complexND} and \eqref{eq:complexNN} with the modes 
shifted as indicated above. For the case of mixed boundary conditions the 
only possible solution is trivial.  The other case, when the pair of real 
coordinates satisfy the same boundary conditions, has  the non-trivial 
solution

\begin{multline}
\label{eq:bsnnt}
\ket{ \eta,g^m}_{\substack{\text{N-N}\\
                                       \text{D-D}}}=
\text{exp}\Big(\mp
\sum_{n=1}^{\infty}\frac{1}{n-m\nu_i}
\beta^i_{-n+m\nu_i}  \wt {\bar \beta}^{j}_{-n+m\nu_i}+
\frac{1}{n+m\nu_i}\bar \beta^i_{-n-m\nu_i} { \wt \beta}^{j}_{-n-m\nu_i}
\\
\mp i\eta\sum_{r>0}^{\infty}(
\lambda^i_{-r+m\nu_i} \wt{\bar \lambda}^{j}_{-r+m\nu_i}+
\bar \lambda^i_{-r-m\nu_i}  \wt { \lambda}^{j}_{-r-m\nu_i})
\Big)\ket{\eta,g^m}^0_{\substack{\text{NS-NS}\\
\text{R-R}}}\;.
\end{multline}
Boundary states on $\IC^3/\IZ_3$ are a tensor product of the boundary states 
along the three complex directions $z^i$. Therefore we conclude that in the 
twisted sector only boundary states with all $s_i$ even exist in this model.

Let us know restore the $(r,{\bf s})$ notation. The twisted sector has zero-modes on the $(x^1,x^2)$ directions. Therefore 
the condition of GSO and orbifold invariance will restrict only the values 
of $r$.
The gluing conditions in the R-R twisted sector are

\begin{align}
\psi^{\mu}_{\eta}\ket{B(r,{\bf s}),\eta}^0_{\RR,\Tw}&=0 && \mu=1, \ldots, 
r+1\nn\\
\psi^{\nu}_{-\eta}\ket{B(r,{\bf s}),\eta}^0_{\RR,\Tw}&=0 && \nu=r+2, \ldots, 
2\nn\\
\end{align}
The  ground states are defined as

\begin{align}
\ket{B(r,{\bf s}),+}^0_{\RR,\Tw}&=a \prod_{\mu=1}^{r+1}\psi^{\mu}_+ 
\prod_{\nu=r+2}^2 \psi^{\nu}_-\ket{B(r,{\bf s}),-}^0_{\RR,\Tw} \nn\\
\ket{B(r,{\bf s}),-}^0_{\RR,\Tw}&=b \prod_{\mu=1}^{r+1}\psi^{\mu}_- 
\prod_{\nu=r+2}^2 \psi^{\nu}_+\ket{B(r,{\bf s}),+}^0_{\RR,\Tw}
\end{align}
with $a$ and $b$ normalization constants. They are related by  
$b=-\frac{1}{a}$.

The representation of the GSO operator on the zero-modes is

\begin{align}
(-1)^F&=\pm 2i \psi^1_0 \psi^2_0  = \pm i (\psi^1_+ + \psi^1_-)(\psi^2_+ + 
\psi^2_-)\nn\\
(-1)^{\wt F} &=\pm 2i {\wt \psi}^1_0  {\wt \psi}^2_0 =\mp i (\psi^1_+ - 
\psi^1_-)(\psi^2_+ - \psi^2_-)
\end{align}
where the phases are determined by the conditions $(-1)^{2{\wt F}}=(-1)^{2F}=1$. The action of these operator onto the boundary states is

\begin{align}
\label{eq:gso}
(-1)^F\ket{B(r,{\bf s}),+}&= \mp i a \ket{B(r,{\bf s}),-}\nn\\
(-1)^F\ket{B(r,{\bf s}),-}&= \mp i b \ket{B(r,{\bf s}),+}\nn\\
(-1)^{\wt F}\ket{B(r,{\bf s}),+}&= \pm (-1)^{r+1}ia \ket{B(r,{\bf 
s}),-}\nn\\
(-1)^{\wt F}\ket{B(r,{\bf s}),-}&= \pm (-1)^{r+1}ib \ket{B(r,{\bf s}),+}
\end{align}
The combination $\ket{B(r,{\bf s}),+}+\ket{B(r,{\bf s}),-}$ will be $(-1)^F$ 
invariant if  $a=b=\pm i$.
The action of $(-1)^{\wt F}$ on this linear combination has eigenvalues 
$\kappa (-1)^{r+1}$, where $\kappa=\pm$.
If we consider that at least one fractional brane, which couples to the 
untwisted and twisted sectors, exists then we have to fix $\kappa = +$.

For IIB, $(-1)^F=(-1)^{\wt F}$  while for IIA $(-1)^F  =-(-1)^{\wt F}$. From 
the last two equations in (\ref{eq:gso}) and the values for $a$ and $b$ one 
finds that  the linear combination   of boundary states in the NS-NS twisted sector with opposite spin structure  is GSO invariant if $r$ is odd in IIB or 
$r$ even in IIA.

Since the twisted  R-R does not have zero-modes along the direction on which 
the orbifold acts, there are no further restrictions on  the twisted R-R 
boundary states.

If we consider that at least one fractional brane, which couples to the 
untwisted and twisted sectors, exists then we have to fix $\kappa = +$.

For IIB, $(-1)^F=(-1)^{\wt F}$  while for IIA $(-1)^F  =-(-1)^{\wt F}$. From 
the last two equations in (\ref{eq:gso}) and the values for $a$ and $b$ one 
finds that  the linear combination is GSO invariant if $r$ is odd in IIB or 
$r$ even in IIA.

Since the twisted  R-R does not have zero-modes along the direction on which 
the orbifold acts, there are no further restrictions on  the twisted R-R 
boundary states. Putting all together one has the results in Table 
\ref{tab:invstatesZ3}.

\section{Mathematical material}\label{sec:mathappendix}

In this paper, we use the fact that the $K$-theory
with compact supports of a space $X$ (satisfying some minimal
topological assumptions which are true here)
is equal to the reduced $K$-theory of the $1$ point compactification
$X^c$ of $X$: formulaically, we write
$$K_{G}^{i,c}(X)=\tilde{K}_{G}^{i}(X^c).$$
Recall that a reduced $G$-equivariant generalized cohomology theory
(such as equivariant $K$-theory)
is applied to a {\em based} $G$-space $X$, which means a space
with a distinguished fixed point called the base point, and usually
denoted by $*$. It is useful to recall also the topological
operation of ``smash product'' of based spaces $X,Y$:
$$X\wedge Y=(X\times Y)/((X\times \{*\})\cup (\{*\}\times Y)).$$
Here we use the topological operation of {\em quotient}, which
means that all the points in the set following the $/$ sign
are identified to a single point. It is worthwhile to note
that the operation $\wedge$ is a commutative associative
unital operation, whose unit is $S^0$, the
$2$-point (fixed) set with one point chosen as base point. Now we have
$$(X\times Y)^c= X^c\wedge Y^c.$$
It is useful for a $G$-space $X$ to denote
$$X_+=X\amalg \{*\}.$$
This is in general a different construction than the
$1$-point compactification $X^c$, although
they are equal when $X$ is compact.
One has
$$K_{G}^{i}(X)=\tilde{K}_{G}^{i}(X_{+}),$$
and when $Y$ is based,
$$\tilde{K}_{G}^{i}(Y)=K_{G}^{i}(Y,\{*\}).$$
For a $G$-representation $V$, the $1$-point compactification
$V^c$ is often denoted by $S^V$.

\vspace{3mm}
A based map
\beg{esf}{f:X\r Y}
is a map of based spaces such that
$f(*)=*$. The {\em based mapping cone} of \rref{esf} is
the based space
\beg{esc}{Cf= (Y\amalg (X\times [0,1]))/(x,1)\sim f(x),
(x,0)\sim (*,t) \;\text{for $x\in X$, $t\in [0,1]$.}
}
The notation after the $/$ sign means that we pass to equivalence
classes of the smallest equivalence relation $\sim$ containing
the relation specified. There is a canonical inclusion $Y\r Cf$,
and one often refers to the sequence
\beg{esc1}{\protect\diagram X\rto^f & Y\rto &Cf
\enddiagram
}
as a {\em cofibration sequence}. The point of considering such
sequences is that they lead to long exact sequences in reduced (equivariant)
generalized cohomology. For example, we have
\beg{esc2}{\r \tilde{K}^{i}_{G}(Cf)\r \tilde{K}^{i}_{G}(Y)
\r K^{i}_{G}(X)\r \tilde{K}^{i+1}_{G}(Cf)\r ...
}

Recall that by a basic property of equivariant generalized cohomology
(sometimes referred to as the Wirthm\"uller isomorphism), for
a subgroup $H\subset G$, we always have
\beg{es5b}{\tilde{K}_{G}^{i}(G/H_+ \wedge X) \cong
  \tilde{K}_{H}^{i}(X),}

notation.
Now for a based space $X$, $S^n\wedge X$ is the (based) $n$-fold
suspension of $X$, so we have, by a general property of generalized
cohomology theories,
\beg{es9}{\tilde{K}^{\ell}_{G}(S^n\wedge X)=
\tilde{K}^{\ell-n}_{G}(X).
}
Because of this, one generalizes this notation
to finite-dimensional dimensional real
$G$-representations $V$ as follows:
\beg{es10}{\tilde{K}^{\ell-V}_{G}(X)=\tilde{K}^{\ell}_{G}(S^V\wedge X)
}
(where $S^V$ was defined above). Now equivariant Bott periodicity
asserts that the sign in \rref{es10} does not matter:
\beg{es11}{\tilde{K}^{\ell-2V}_{G}(X)=\tilde{K}^{\ell}_{G}(X).
}
We can therefore take as ``dimension'' of the equivariant
$K$-theory group any element of the real representation ring
$RO(G)$. It is worthwhile to note that we essentially now reviewed
the entire definition of an equivariant cohomology
theory: the basic properties
are the stability under suspension \rref{es10}, the long exact
sequence \rref{esc2}, and the indexing by elements of $RO(G)$;
the last property is sometimes deleted or modified, but it holds
for $K$-theory. For details on equivariant stable homotopy
theory, we refer the reader to \cite{lms}.

\vspace{3mm}
In fact, however, in the case of equivariant $K$-theory,
by equivariant Bott periodicity, a simplification occurs.
Dimensions belonging to the complex representation ring $R(G)$
can be identified with $0$, and the group of non-trivial
dimensions is
\beg{es12}{D(G)=RO(G)/R(G).
}
In \rref{es12}, the embedding $R(G)\subset RO(G)$ takes
a complex representation to the underlying real representation.
This is not a map of rings; rather, the image is an ideal of $RO(G)$.


There are two basic elementary principles which aid us in the
calculation. First of all, assume that the representation $V$ is
trivial when restricted to some subgroup $A\subset G$. Then, by a general
principle of equivariant cohomology sometimes referred to as
the Adams isomorphism,
\beg{es17}{\tilde{K}^{i}_{G}(S^V)\cong \tilde{K}^{i}_{G/A}(S^V)\otimes R(A)
\cong \tilde{K}^{i}_{G/A}(S^V)\otimes \Z^{|A|}.
}
The other principle is that when $V=\gamma_1\oplus...\oplus \gamma_k$
where $\gamma_j$ are $1$-dimensional real representations independent
in the character group, then $\epsilon=0$ and
\beg{es18}{\tilde{K}^{0}_{G}(S^V)\cong \Z^{2^{n-k}}.
}
To see this, one just takes the smash product of cofiber sequences
of the form
\beg{es19}{G/Ker(\gamma_j)_+\r S^0\r S^{\gamma_j}
}
(where for a representation $\gamma$, $Ker(\gamma)$ is the maximal subgroup
restricted to which $\gamma$ becomes trivial), using
\rref{es5b}.

\end{document}